\documentclass[runningheads]{llncs}
\usepackage{amsmath}
\usepackage{booktabs} 
\usepackage{caption} 
\usepackage{graphicx}
\usepackage{pgfplots}
\usepackage[all]{nowidow}
\usepackage[utf8]{inputenc}
\usepackage{tikz}
\usetikzlibrary{er,positioning,bayesnet}
\usepackage{multicol}
\usepackage{algpseudocode,algorithm,algorithmicx}
\usepackage[left=2.54cm,right=2.54cm,bottom=2.54cm,top=2.54cm]{geometry}
\usepackage{setspace}
\usepackage{multirow}
\usepackage[superscript,biblabel]{cite}

\usepackage{titlecaps}
\Addlcwords{the of into by for from in on to and a an at with as around after along or nor but yet so all via 1-nm its above using based under below through considering}
\definecolor{blue}{HTML}{1F77B4}
\definecolor{orange}{HTML}{FF7F0E}
\definecolor{green}{HTML}{2CA02C}

\pgfplotsset{compat=1.14}

\begin{document}
\title{High Temperature Thermal Photonics}
%
%
\author{Xueji Wang$^{1}$, Ryan Starko-Bowes$^{2}$, Chinmay Khandekar$^{1}$ and Zubin Jacob*$^{1}$}
%
%
\institute{$^{1}$Birck Nanotechnology Center, Purdue University, West Lafayette, Indiana 47907, USA
\\$^{2}$University of Alberta, Edmonton, Alberta T6G 2R3, Canada
\email{\\$^{*}$zjacob@purdue.edu}\\}
\maketitle              
\begin{abstract}
Controlling and detecting thermal radiation is of vital importance for varied applications ranging from energy conversion systems and nanoscale information processing devices to infrared - imaging, spectroscopy and sensing. We review the field of high temperature thermal photonics which aims to control the spectrum, polarization, tunability, switchability and directionality of heat radiation from engineered materials in extreme environments. We summarize the candidate materials which are being pursued by the community that have simultaneous polaritonic/plasmonic properties as well as high temperature stability.  We also provide a detailed discussion of the common photonic platforms including meta-gratings, photonic crystals, and metamaterials used for thermal emission engineering. We review broad applications including thermophotovoltaics, high temperature radiative cooling, thermal radiation sources, and noisy nanoscale thermal devices. By providing an overview of the recent achievements in this field, we hope this review can accelerate progress to overcome major outstanding problems in modern thermal engineering.

\end{abstract}
\section{Introduction}


The field of high temperature thermal photonics deals with the control of radiative heat by nanostructures, metamaterials, photonic crystals and engineered materials (Table \ref{table:structures}). The key properties of radiation excited by thermal agitation include the spectrum (temporal coherence), polarization (spin), directivity (spatial coherence) and total energy conversion efficiency. The black-body thermal radiation is spectrally peaked at room temperature in the mid-infrared spectrum often called the long-wave infrared (LWIR) region, whereas at high temperature the thermal energy density shifts to the near-IR and visible spectrum. This broad spectral change with temperature poses unique challenges for arbitrary control over thermal radiation. This endeavor requires intertwined strides in fundamental radiative heat transfer theory, nanomaterials development with desired properties and simultaneous robustness, computational design of optimal nanostructures and high precision extreme environment thermal metrology.

\definecolor{Gray}{gray}{0.9}
\begin{table}[!htb]
\caption{Summary of common photonic structures in high temperature thermal emission engineering}
\centering
\setlength{\tabcolsep}{4mm}{ 
\begin{tabular}{ ccc }
\toprule
& Strucutures & References \\
\hline
\multirow{3}*{Gratings}
& 1D Gratings & \cite{SiC_1DGrating,Tungsten_Grating2} \\
& 2D Gratings & \cite{Greffet2D,RyanSiCGratings} \\
& Bullseye Gratings & \cite{Source_Bulleye,Source_BulleyeTh}  \\
\hline
\multirow{3}*{Metasurfaces} 
& Plasmonic Metasurfaces & \cite{MM_SANTE,MM_TETung} \\
& Polaritonic Metasurfaces & \cite{MM_Refr,MM_Refr2} \\
& Semiconductor Metasurfaces& \cite{MM_SemiSirods,MM_Refr5} \\
\hline
\multirow{3}*{ENZ Materials}
& ENZ coatings & \cite{ENZ_Coating2,ENZ_Coating} \\
& ENZ/Hyperbolic Metamaterials & \cite{ENZ_MMENP,ENZ_MMExp} \\
& Arbitrarily shaped ENZ bodies & \cite{ENZ_Body} \\
\hline
\multirow{4}*{Photonic Crystals}
& 1D Multilayer Stacks & \cite{TPV_ExpW2.5,TPV_SExp3.2} \\
& 2D Air-hole Arrays & \cite{PC_2DCore,PC_2DTa2} \\
& 3D Woodpile Structures  & \cite{PC_3DWoodPile1,PC_3DWoodPile3Coating} \\
& 3D Inverse-opal Structures & \cite{PC_3DInverseM1,PC_3DInverse3NC} \\
\hline
\multirow{3}*{Nanotube Systems}
& CNT Films/Bundles & \cite{CNTBB_VASWCNT,CNTEmitter_HighSpeed} \\
& Individual CNTs & \cite{CNTBB_IndividualMWCNT,CNTBB_IndividualSWCNT}\\
& BNNTs & \cite{BNNT_Ryan} \\
\hline
\multirow{2}*{Others}
& Quantum Wells & \cite{NodaMQW_NP,NodaMQW_NM} \\
& Cold-side Filtering & \cite{Source_NN,Source_Fan} \\
\bottomrule
\end{tabular}}
\label{table:structures}
\end{table}


The theoretical foundations of the field are connected firstly to Planck's law of black-body radiation\cite{PlanckFirst}. This specifically connects the spectrum of heat radiation to the Bose-Einstein distribution of photons and the quantum of energy in a bosonic harmonic oscillator. For practical engineering applications, researchers appeal to Kirchhoff's laws which states that the angle-resolved thermal emission spectrum can be designed directly by engineering the angular absorption spectrum\cite{KirchhoffFirst}. Important theoretical frontiers are related to generalizing Kirchhoff's laws for non-equilibrium\cite{FanNonEquil,NonEqu_Greffet,ChinmaySpin1} and non-reciprocal\cite{FanNonRecip,ChinmaySpin2} systems and also quantifying properties like spin\cite{ChinmaySpin1,ChinmaySpin2} which were overlooked previously. The near-field radiative heat transfer requires fluctuational electrodynamics theory, fundamentally beyond conventional Kirchhoff's laws. This technique pioneered by Rytov rests on the fluctuation-dissipation theorem \cite{RytovFirst} also known in different research communities as the Kubo formalism, Callen-Welton theorem and Johnson-Nyqvist noise. The most striking physical phenomenon predicted by Rytov's theory is super-Planckian thermal emission, which is connected with the ability of polaritonic materials and engineered metamaterials to enhance radiative heat transfer beyond Planck's far-field blackbody limit. 

~\begin{figure}
    \centering
    \includegraphics[width=0.7\columnwidth]{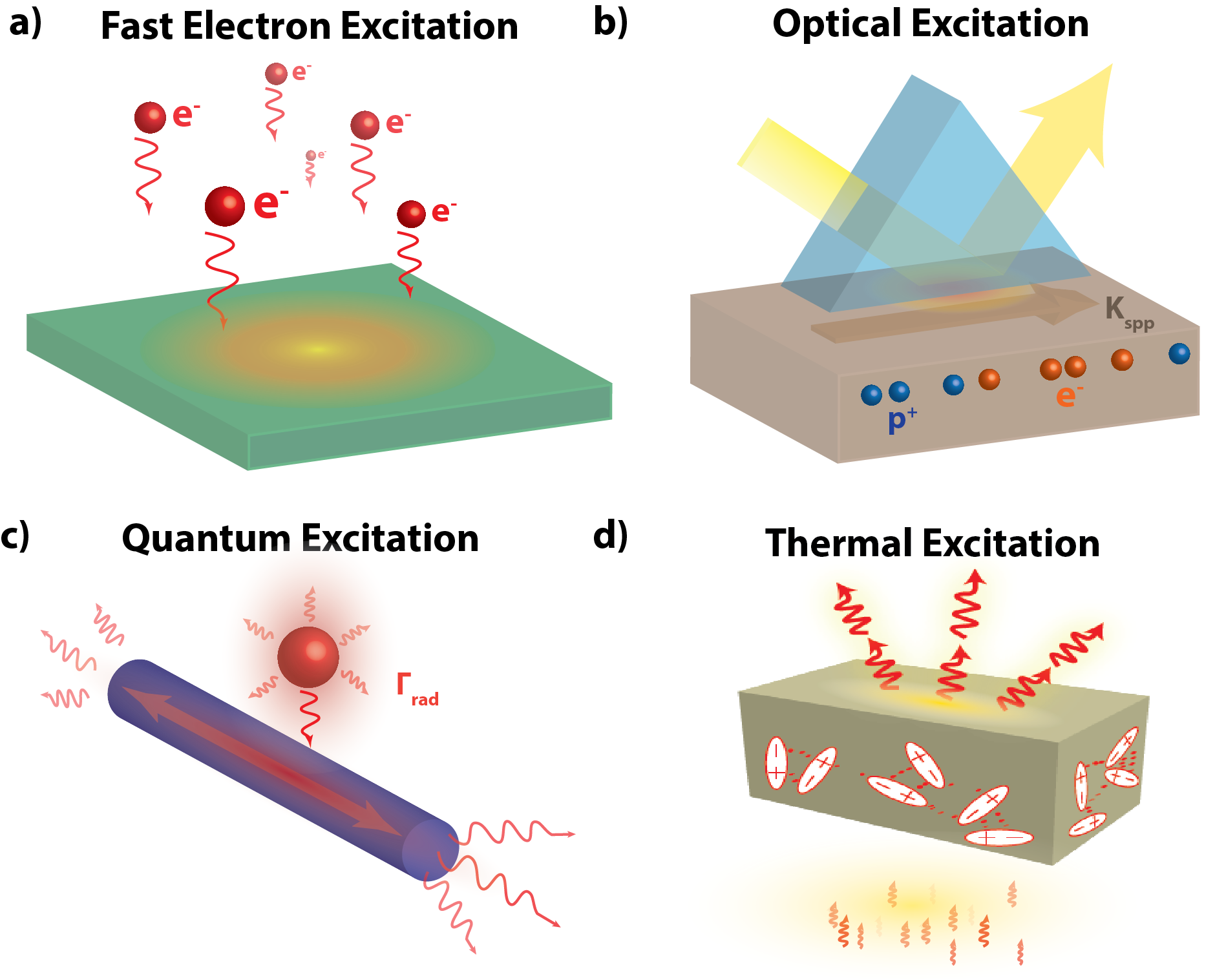}
    \caption{Excitation methods for plasmon/phonon polaritons. a) An incident free electron beam induces a plasmon excitation in the material. b) Evanescent waves which can be generated via total reflection at the bottom interface of a prism, excites surface plasmons by directly coupling. c) A single quantum dot excites surface plasmons by directly coupling to the plasmonic waveguide. d) Heating plasmonic/polaritonic materials to high temperatures thermally excites the plasmon/phonon polaritons.}
    \label{fig:excitations}
\end{figure}

Fig. 1 shows the central concept of thermally excited polaritons and contrasts it from other research areas of light-matter interaction. The plasmon-polariton which is a collective oscillation of the free electrons of a metal with light was first discovered (Fig. 1(a)) by excitation using fast-electrons\cite{PlasmonFirst}. Subsequently, optical laser excitation (Fig. 1(b)) using unique momentum matching techniques was made possible for applications such as nanoscale biosensing\cite{KretschmannFirst,OttoFirst}. The modern techniques of coupling to surface plasmon polaritons (SPPs) is quantum optical in nature (Fig. 1(c)) where a single quantum emitter spontaneously emits a single plasmonic mode\cite{Lukin_QuanExcitation}. The recent frontier in the field is related to thermal excitation (Fig. 1(d)) of phonon-polaritons and plasmon-polaritons for high temperature applications.


One major challenge for the field is related to the development of materials which possess desired optical/IR properties and simultaneously exhibit stability to extreme thermal environments. The optical polarizability and absorption losses are intimately connected to the thermal emission but are also functions of temperature. A systematic classification and categorization of high temperature optical effects are underway to gain control over microscopic mechanisms that affect key figures of merit. In this regard, thermal photonics is a major application driver for the field of plasmonics and metamaterials since absorption losses are no more detrimental but are in fact necessary for optimal engineering. We review in detail the candidate materials which are being pursued by the community including polaritonic dielectrics, high-temperature metals, intrinsic/doped semiconductors and other refractory plasmonics (Table \ref{table:materials}). 

\definecolor{Gray}{gray}{0.9}
\begin{table}[!htb]
\caption{Summary of common refractory materials in high temperature thermal photonics}
\centering
\setlength{\tabcolsep}{4mm}{ 
\begin{tabular}{ ccc }
\toprule
 & Materials & Characteristics\\
\hline
\multirow{4}*{Polaritonic Dielectrics} 
              & SiC & Low optical loss \\
              & & Superior thermal stability\\
               \cline{2-3}
              & hBN &  Natural hyperbolic material\\
              & & Highest optical phonon frequencies\\
\hline
\multirow{7}*{Plasmonic Materials} 
             & Refractory Metals & Relatively high loss \\
             & (W, Pt, Ta) &Weak plasmonic response\\
              \cline{2-3}
             & Intrinsic/Doped Semiconductors &  Tunable optical properties  \\
             & (Si, III-V semiconductors, TCOs) &  Mature manufacturing techniques \\
             \cline{2-3}
		     & Transition Metal Nitrides & Good thermal stability \\
		     & (TiN, TaN, ZrN, HfN) & Strong plasmonic response \\
		     \cline{2-3}
		     & Graphene &   Large tunability \\
		     & & Superior  thermal  conductivity\\
\bottomrule
\end{tabular}}
\label{table:materials}
\end{table}

As the scope of naturally occurring materials is limited, metamaterials, artificial materials that can be engineered to have desired optical properties at desired wavelength ranges, are of interest to the thermal engineering community\cite{MM_Book1,MM_Book2}. In the area of high temperature thermal photonics, these metamaterials rest on refractory building blocks and generate an array of unique thermal emission patterns. Here, we review the high temperature metamaterials that are based on various electromagnetic resonances in natural refractory materials, including plasmon-polariton resonances, phonon-polariton resonances, and interband transitions. We also highlight the advantages of epsilon-near-zero (ENZ) materials in tailoring high-temperature thermal emission.   We show that metamaterials with giant anisotropy provide another unique approach to thermal emission engineering. Different dielectric responses along different optical axes can be used to achieve spatially distinct and independently controllable thermal emission patterns. We also review the unique characteristics of hyperbolic metamaterials, where the nontrivial hyperbolic topologies of photonic isofrequency surfaces result in the super-Planckian near-field radiative heat transfer as well as strongly suppressed far-field thermal radiation.

Photonic crystals (PhCs), with the characteristic feature of photonic bandgaps, are also widely used as high temperature thermal emitters. As discussed in a subsequent section, strongly suppressed thermal emission has been demonstrated within the bandgaps, while enhancement can be realized by quality factor matching (Q-matching). We review the design principles and experimental progress of metallic 2D PhCs and introduce the inverse opal 3D photonic crystals. We discuss their advantages in the application of high temperature thermal photonics compared with other photonic designs.

Additionally, we show that nanotube systems can also be exploited in thermal emission engineering due to their unique 1D structures. We review the various applications of different carbon nanotube (CNT) systems including ideal blackbody sources and high-speed on-chip emitters. We also show that the tubular form of the hexagonal boron nitride (hBN): boron nitride nanotubes (BNNTs) can work as natural polaritonic thermal antennas with striking high temperature stability.


Fig. 2 depicts one of the outstanding challenges for the field of high temperature thermal photonics. It shows a highly efficient ‘polaritonic globar’. Traditional globars are widely used in industry as infrared spectroscopy sources and have current driven broadband thermal emission. The globar is often made of silicon carbide (SiC) composites with flat emissivity spectra. One long term goal for the field is not only to enhance efficiency of current-to-heat conversion but also develop a multi-functional modern globar with  spectral as well as  polarization control, tunability, switchability, and beam directionality. 

~\begin{figure}[!ht]
    \centering
    \includegraphics[width=0.45\columnwidth]{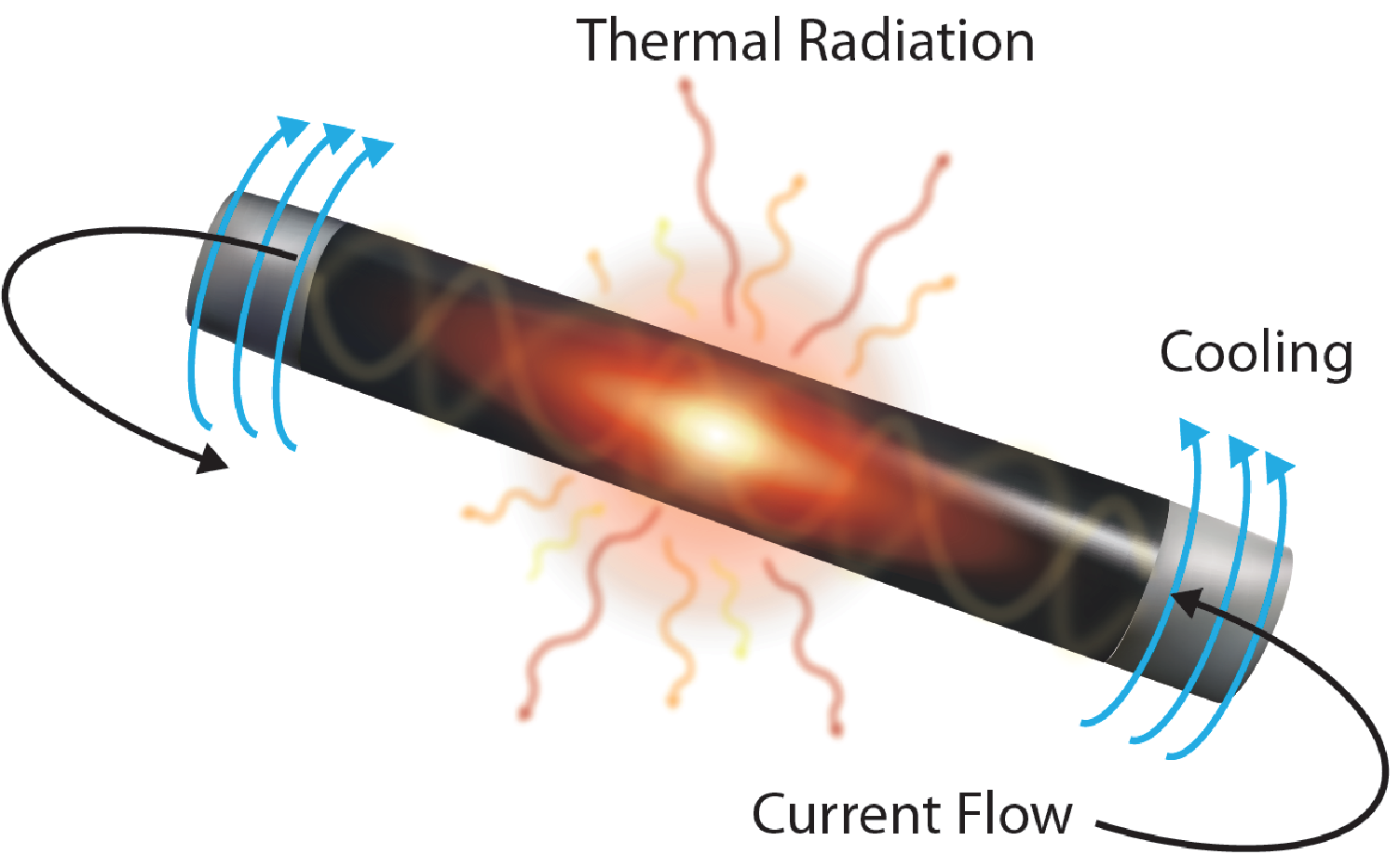}
    \caption{ Schematic of a 'Globar' light source.}
    \label{fig:Globar}
\end{figure}


The fast development of high temperature thermal photonics benefits many industrial and scientific applications. Here, we also review the recent achievements of high-efficiency thermophotovoltaics (TPV) systems, which are enabled by the efficient solar absorbers and selective emitters based on a variety of refractory photonic designs including PhCs and metamaterials. We discuss the experimental demonstrations of high-speed, highly directional and narrow-band high temperature thermal sources for broad applications including gas sensing, illumination, and display. We highlight that these thermal sources can be integrated into on-chip systems to achieve compact functional devices for commercialization. We discuss the fundamentals and developments of radiative cooling and show that this approach, which is initially proposed and experimentally demonstrated at room temperatures, can be extended to the high-temperature regime, which is of vital importance for some extreme environments from industrial manufacturing to space exploration.


The review is arranged as follows. We first provide an exhaustive list of properties of high temperature materials. This is followed by two sections on high temperature metamaterials and ENZ materials where the focus is on design methodologies to control far-field thermal emission. The next section deals with the concept of thermal photonic crystals where periodic microstructuring leads to controlled emission of thermally excited radiation. We then go into details of various classes of nanotube emitters in section 6. Finally, we discuss multiple application drivers and emerging frontiers for the field of high temperature thermal photonics. 

\section{High-Temperature Materials}

\subsection{Polaritonic Dielectrics}

Polaritonic dielectric materials are the most important class of materials for the field of thermal photonics. Dielectric materials where heat conduction is dominated by phonons are often transparent and insulating. However, there is a class of dielectric materials with optically active phonons which arises from a polar response i.e. a nonvanishing dipole moment between constitutent elements. This causes a screening of electromagnetic fields i.e. metallic behavior in a specific spectral window known as the Reststrahlen band. In this region between transverse (TO) and longitudinal optical (LO) phonons, electromagnetic waves strongly couple with the vibrational motions of the polar lattice leading to phonon polaritons, coupled oscillations of light and matter. Thus, thermally exciting such materials leads to the strong fluctuating optical dipole moments and possibility to control heat radiation in well-defined spectral regions. An important thermal excitation in such materials is the surface phonon polaritons (SPhPs), which is an electromagnetic wave in the Reststrahelen band that travels on the interface of a polaritonic dielectric and vacuum.

~\begin{figure}[!ht]
    \centering
    \includegraphics[width=0.7\columnwidth]{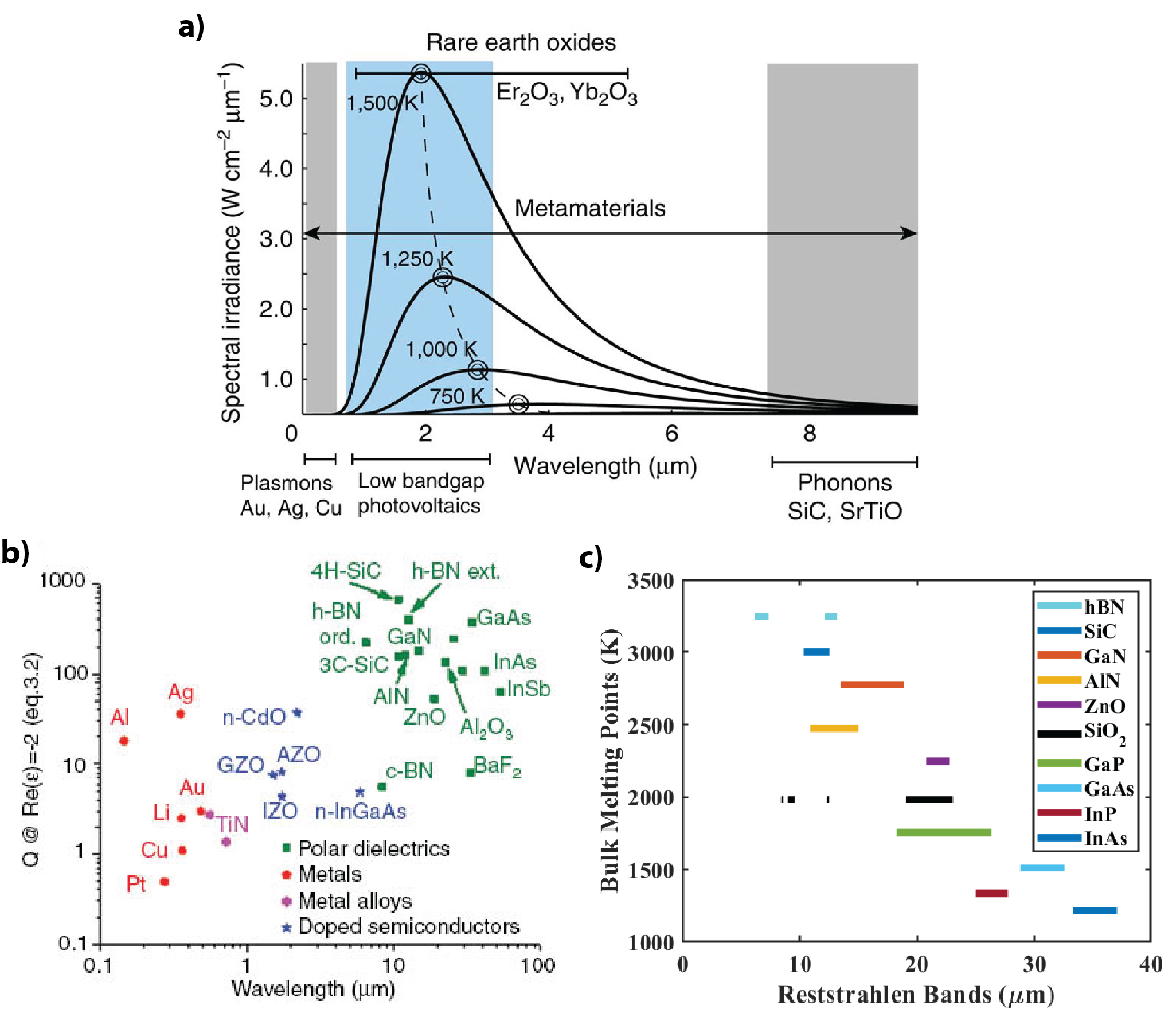}
    \caption{Material choices for high temperature thermal photonics. a) Overlap of the spectral irradiance of a blackbody half-space with natural optical resonances. Adapted from Ref. \cite{ENZ_MMExp}. b) Comparison of quality factor (Q) for a spherical particle in air for plasmonic materials (metals, metal alloys and doped semiconductors) and polaritonic materials (polar dielectrics). Plasmonic materials show higher operating frequencies but higher losses, while polaritonic materials possess lower losses but the operating frequency is also reduced by an order of magnitude. Adapted from Ref. \cite{Polar_LowLoss}. c) Reststrahlen bands and melting points of different polar dielectrics. Bulk hBN and SiC show both high melting points and high phonon frequencies, which make them appealing for high temperature thermal engineering.}
    \label{fig:Materials}
\end{figure}

To understand the importance of this class of materials for high temperature thermal photonics, we study Fig. \ref{fig:Materials}(a), which shows the overlap  of  the  spectral  irradiance  of a  blackbody  half-space  with  natural  optical  resonances \cite{ENZ_MMExp}. An important contrast is revealed from conventional plasmonic metals where the bulk plasmon frequency $\omega_p$ is proportional to 
$(N/m_e)^{1/2}$ \cite{ENZ_Plasma1}. Here, $m_e$ and $N$ are the effective mass and concentration of electrons, respectively. The  relevant resonant frequency lies in the visible to near-infrared range due to the small value of effective electron mass and high electron concentration of typical plasmonic materials. In polaritonic dielectrics, the resonant frequency of LO ($\omega_{LO}$) and TO phonons ($\omega_{TO}$) are inversely proportional to the square root of the oscillator mass ($\omega_{LO},\omega_{TO}\propto(1/m_a)^{1/2}$) \cite{ENZ_Plasma1}. Due to the large mass of the oscillating ions, the frequencies always lie in the mid-infrared (Table \ref{table:hBNfrequencies}). Thus the overlap with the blackbody spectrum is substantial paving the route for efficient thermal engineering. Another major difference between polar dielectrics and plasmonic materials is the optical loss. Plasmon polaritons are the collective oscillations of free electrons, where the high concentration of electrons and a relatively fast electron scattering time make plasmonic materials more lossy than polaritonic dielectrics \cite{Polar_Scattering}. While a small amount of absorption loss is necessary for thermal emission due to Kirchhoff's laws, a large loss leads to difficulty in achieving high quality resonances.

\definecolor{Gray}{gray}{0.9}
\begin{table}[!htb]
\caption[Optical phonon frequencies of hBN and other common polaritonic materials.]{Optical phonon frequencies of common phonon polaritonic materials. The in plane optical phonons of hBN have the highest frequencies. Adapted with permission from Ref. \cite{BNNT_Ryan}. Copyright 2019, American Chemical Society}
\centering
\setlength{\tabcolsep}{4mm}{ 
\begin{tabular}{ cccccc }
\toprule
Material & $\omega_{LO}$ $(\lambda_{LO})$ & $\omega_{TO}$ $(\lambda_{TO})$ & $\gamma$ & $\varepsilon_\infty$ & Ref.\\
Permittivity & [cm$^{-1}$] ($\mu$m) & [cm$^{-1}$] ($\mu$m) & [cm$^{-1}$] & &\\
\hline
hBN ($\varepsilon_\parallel$/$\varepsilon_{t,a}$) & 1614 (6.20) & 1360 (7.35) & 7 & 4.90 & \cite{CaldwellSubDiffhBN} \\
hBN ($\varepsilon_\perp$/$\varepsilon_{r}$) & 825 (12.12) & 760 (13.16) & 2 & 6.76 & \cite{CaldwellSubDiffhBN} \\
\hline
cBN & 1340 (7.46) & 1065 (9.39) & 40.5 & 4.5 & \cite{Gielisse:cBNOptPhon} \\
\hline
 & 1186 (8.43) & 1167 (8.57) & 4.43 & &  \\
 & 1112 (9.00) & 1046 (9.56) & 15.53 & &  \\
SiO$_2$ & 1106 (9.04) & 1058 (9.45) & 0.42 & 2.09 & \cite{KischkatSiO2TF} \\
 & 814 (12.29) & 799 (12.52) & 12.94 & &  \\
 & 525 (19.05) & 434 (23.04) & 54.14 & &  \\
\hline
SiC & 969 (10.32) & 793 (12.61) & 4.76 & 6.70 & \cite{Greffet2D} \\
\hline
 & 788 (12.69) & 543 (18.42) & 17.0 & &  \\
SrTiO$_3$ & 474 (21.10) & 175 (57.14) & 5.4 & 5.10 & \cite{VogtSrTO3OptPhon,KamarasSrTO3OptPhon,GERVAISSrTiO3OptPhon} \\
 & 172 (58.14) & 91 (109.89) & 15.0 & &  \\
 \hline
 $\alpha$-GaN & 740 (13.51) & 530 (18.87) & 7 & 5.40 & \cite{ShkerdinGaNOptPhon} \\
 $\beta$-GaN & 739 (13.53) & 553 (18.08) & 7 & 5.35 & \cite{ShkerdinGaNOptPhon} \\
 \hline
 GaP & 403 (24.8) & 367 (27.25) & 1.29 & 9.09 & \cite{FischerGaPOptPhon} \\
\bottomrule
\end{tabular}}

\label{table:hBNfrequencies}
\end{table}

Fig. \ref{fig:Materials}(b) shows the quality factor Q of the polaritonic (plasmonic) mode in polar dielectrics (plasmonic materials) versus the corresponding wavelength ($Q_{Re(\epsilon)=-2}$ versus $\lambda_{Re(\epsilon)=-2}$) \cite{Polar_LowLoss}. In contrast to plasmonic materials, polar dielectrics show both higher quality factors (lower loss) and longer resonance wavelengths located in the mid-infrared range. Additionally, Fig. \ref{fig:Materials}(c) shows the Reststrahlen bands and melting points of bulk polaritonic dielectrics. Many polaritonic dielectrics possess higher thermal stability compared with these conventional plasmonic materials such as Au and Ag. These advantages make polaritonic dielectrics a unique platform for engineering high-temperature thermal radiation. 

\subsubsection{Silicon Carbide}

Among all polaritonic dielectrics, SiC is one of the most commonly used materials exploited for high-temperature thermal photonic systems. Fig. \ref{fig:Materials}(b) and Fig. \ref{fig:Materials}(c) reveal that SiC retains both low optical loss and superior thermal stability. SPhPs that exist on the surface of a SiC plane can be thermally excited at high temperatures. In 2002, a pioneering work  by J. J. Greffet et al. showed that coherent far-field emission from a thermal source was possible by coupling the SPhPs to free space radiation via surface relief gratings in SiC \cite{SiC_1DGrating}. Many subsequent important demonstrations in SiC followed, such as selective thermal emitters based on individual SiC polaritonic antennas \cite{SiC_Antenna} and metasurfaces \cite{MM_Refr2}. Recently, Starko-Bowes et al. demonstrated a dual band quasi-coherent thermal source based on a 2D SiC grating device \cite{RyanSiCGratings}. Two distinct emission bands with different polarization are produced at high operating temperatures up to 963K. 

\subsubsection{Hexagonal Boron Nitride}

hBN has the same hexagonal crystal structure as graphene but with alternating boron and nitrogen atoms in place of carbon. Unlike graphene, hBN is a polaritonic dielectric (insulating) material. Since both nitrogen and boron are light elements, hBN has the highest optical phonon frequencies among known polar dielectrics as shown in Table \ref{table:hBNfrequencies}. By rolling hBN sheets into a tubular geometry, BNNTs with one-dimensional structure similar to CNTs are formed. \cite{BNNT_First,BNNT_Syn,BNNT_Stability,BNNT_RecentReview1,BNNT_RecentReview2}. Both bulk hBN and BNNTs possess good thermal stability, which make them appealing for high temperature thermal engineering. These class of materials can be combined with existing industrial standards of ceramics to create the next generation of 'polaritonic ceramics' for commercial applications.

\subsection{Plasmonic Materials}

The collective and coherent quasi-particle oscillation of free electrons in matter, known as a plasmon, is a key enabling phenomenon in nanophotonics \cite{Plasmonics_Book}. Noble metals such as gold and silver, are the most commonly studied materials due to the relatively low loss among plasmonic materials. However, their high cost, low mechanical robustness and high loss of plasmonics in general largely limit their usage in practical applications. In particular, the poor thermal stability of gold and silver exclude them from applications for high-temperature photonics. This was pointed out by S. Molesky along with the first proposal to use alternative plasmonic materials for thermal photonics \cite{ENZ_MMENP}.  Plasmonic materials fall in various material classes including metals (Ag, Au, Cu, etc.), heavily doped semiconductors (n-Si, p-Si, n-GaAs, p-GaAs, Al:ZnO, etc.), silicides, germanides, transition metal nitrides (TiN, ZrN, TaN, etc.), and 2D plasmonic materials (Graphene, MoS$_2$, etc.) \cite{Plasmonics_Overview}. Following the initial proposal of high temperature plasmonics by S. Molesky et al., many theoretical and experimental efforts have been made to find low loss plasmonic materials that can survive at sustained high temperatures, known as refractory plasmonics \cite{Plasmonics_Refractory}. Below, we select some typical refractory plasmonic materials and provide a brief overview of their characteristics.

\subsubsection{Refractory Metals}
Noble metals are the most widely used plasmonic materials due to their ease of fabrication and ubiquity in academic labs. Refractory metals such as tungsten (W), tantalum (Ta) and platinum (Pt) are another subset of metallic materials with the advantage that they possess a superior ability to withstand sustained high temperatures \cite{PC_2DCore,PC_2DTa2,Metal_Pt}. Recent works have demonstrated that selective thermal emitters can be built by etching gratings into refractory metal thin films to couple the thermally-excited SPPs with the free space radiation continuum \cite{Tungsten_Grating1,Tungsten_Grating2}.  Additionally, the plasmonic resonance of refractory metals have been used to build refractory metasurfaces that enhance both the temporal and spatial coherence of the thermal emission \cite{MM_Refr3}. W and Ta PhC emitters have also been experimentally demonstrated and show strong enhancement of thermal emission spectra at near infrared wavelengths \cite{PC_2DCore,PC_2DTa1,PC_3DWoodPile1}. Suppressed thermal emission at the long-wavelength range is naturally guaranteed in these PhCs due to the intrinsically low emissivity at frequencies below the plasma frequency. 

However, it should be noted that refractory metals generally have a weak plasmonic response. Another strategy to achieve refractory plasmonics is to use conventional plasmonic materials with high temperature stabilizing coatings. For example, Albrecht et al. demonstrated that by overcoating gold nanostructures with alumina, fabricated devices can withstand temperatures of over 800 $^\circ$C and intense laser radiation of over 10 GW/cm$^2$ at ambient atmospheric conditions \cite{Metal_RefractaryGold}. In this way, the excellent plasmonic properties of gold can be combined with the properties of refractory materials and get exploited in high temperature thermal photonics.

\subsubsection{Heavily Doped Semiconductors}

When semiconductor materials are heavily doped, the extremely high carrier concentration makes them have optical and near-infrared Drude-like responses similar to metals. Additionally, conventional semiconductors such as silicon (Si) and III-V semiconductors such as gallium nitride (GaN) and aluminum nitride (AlN) also simultaneously show excellent thermal stability. Transparent conducting oxides (TCOs) such as indium tin oxide (ITO), aluminum-doped zinc oxide (AZO) and gallium doped zinc oxide (GZO) are also a subset of heavily doped semiconductors with metallic response.

One of the most attractive advantages of doped semiconductors is the ability to tune their optical properties by controlling the dopant (and charge carrier) concentration. Both the plasma frequency and material loss are dependent on the carrier concentration \cite{ENZ_Plasma1}. For III-V semiconductors, optical bandgaps are also tunable by varying the composition of the compounds. Thus the interband transition offers another degree of freedom to get desired optical properties. 

For example, Ribaudo et al. built a quasi-coherent thermal emitter by etching a 2D grating structure into highly doped n-Si with a phosphorous concentration of $2.4$x$10^{20}$ cm$^{-3}$ \cite{Semi_SiGrating}. This produces a plasma frequency in the mid-infrared spectral range. SPPs at the 10-12 $\mu$m range are coupled into free space by the grating structure, leading to highly directional thermal emission. Since the nanofabrication and relevant technologies of conventional semiconductors have been well developed by the semiconductor industry, devices based on these materials provide a promising platform for exploring high-temperature thermal photonics.

\subsubsection{Graphene}

Graphene, single atom-thick layers of carbon, has a tunable plasmon frequency generally in the infrared range. Due to graphene's superior thermal conductivity \cite{Graphene_ThermalCond}, strong light-matter interaction \cite{Graphene_LightInter} and large electrical tunability \cite{Graphene_Tune}, it has also been demonstrated as a promising platform for high temperature plasmonics. Various novel thermal emitter designs have been proposed using graphene. In one study, Pu et al. numerically demonstrated a narrow band and highly directional thermal emitter from a single layer graphene coating \cite{Graphene_Single}. Tunable optical metamaterials with hyperbolic dispersion are also achievable by stacking graphene sheets and dielectric layers \cite{Graphene_Meta1,Graphene_Meta2}. Pendharker et al. calculated the far-field emissivity of the multilayered graphene metamaterials with an effective permittivity approach. Their result suggests that selective thermal emission with strongly suppressed long-wavelength emission and enhanced short-wavelength emission can be achieved in such structures. \cite{ENZ_Coating}

\subsubsection{Transition Metal Nitrides}

The field of high temperature plasmonics and metamaterials necessarily requires new materials since traditional noble metals have poor thermal and mechanical stability. S. Molesky et al. proposed that a major application of alternative plasmonic materials is in high temperature photonics \cite{ENZ_MMENP}. Following this proposal, transition metal nitrides such as titanium nitride (TiN), tantalum nitride (TaN), zirconium nitride (ZrN) and hafnium nitride (HfN) were later demonstrated to be very promising materials for refractory plasmonics. They exhibit good metallic properties similar to other low loss metals (Ag, Au etc.) with plasmonic resonances from visible to NIR range even at high temperatures \cite{Nitrides_TiNRing,TiN_Giessen}. Apart from the excellent thermal stability (bulk melting points around 3000 K), they also show good chemical stability and high mechanical robustness, making them enticing candidates for extreme environment applications. 

~\begin{figure}[!ht]
    \centering
    \includegraphics[width=0.7\columnwidth]{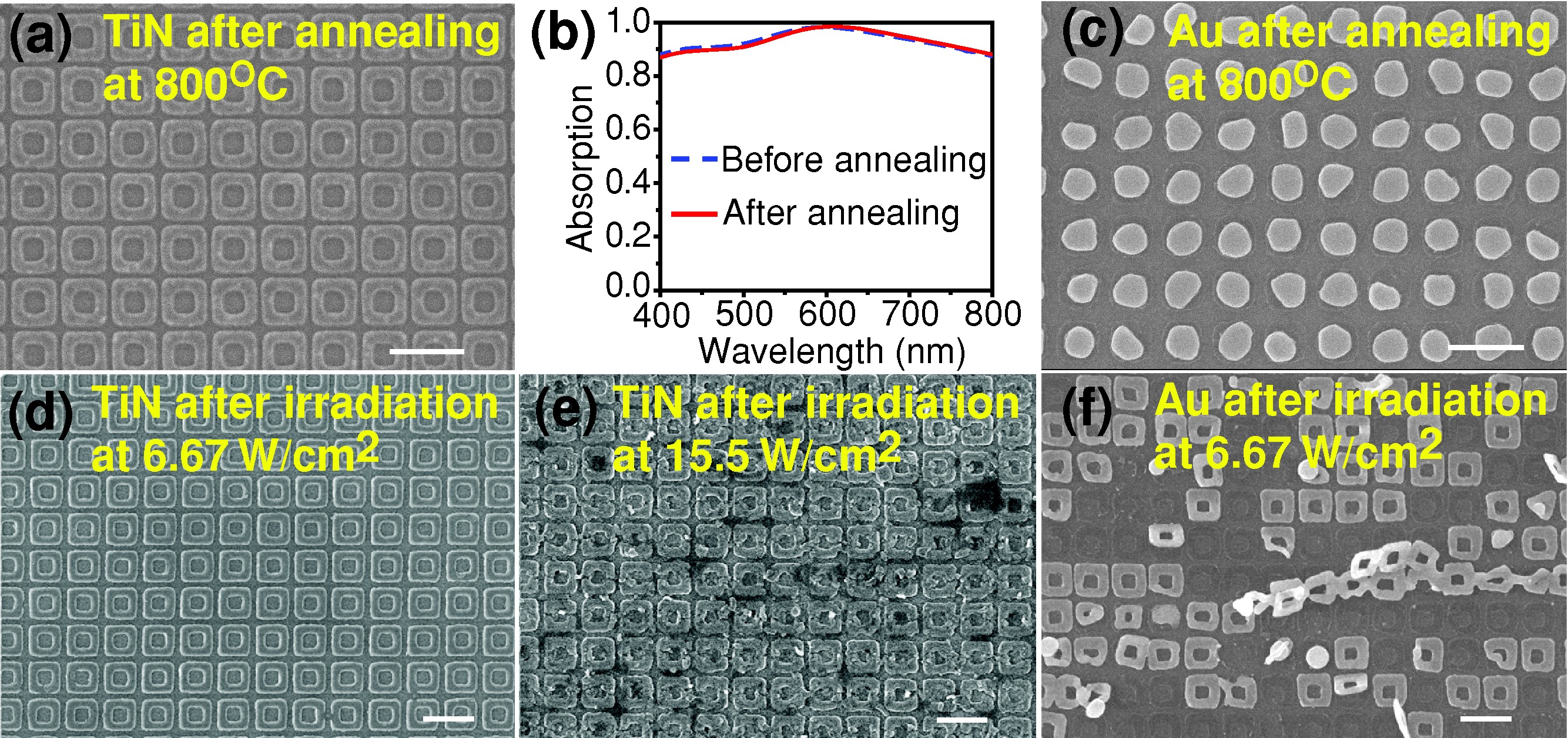}
    \caption{Comparison of the thermal stability for TiN and Au metasurfaces. a) SEM image of the TiN absorber after annealing at 800 $^\circ$C for 8 h. b) The measured absorption spectrum of the TiN absorber after annealing at 800 $^\circ$C for 8 h, compared with the measured absorption spectrum of TiN before annealing. c) SEM image of the Au absorber after annealing at 800 $^\circ$C for 15 min. d,e) SEM image of the TiN absorber after laser excitation with an intensity of 6.67 W/cm$^2$ (d) and 15.5 W/cm$^2$ (e). f) SEM image of the Au absorber after laser excitation with an intensity of 6.67 W/cm$^2$. The scale bars are 400 nm. Adapted with permission from Ref. \cite{Nitrides_TiNRing}. Copyright 2014, WILEY‐VCH Verlag GmbH \& Co. KGaA, Weinheim.}
    \label{fig:TiN}
\end{figure}

As shown in Fig. \ref{fig:TiN}, Li et al. fabricated a TiN ring array solar absorber with an average absorption of 95\% over the spectral range of 400–800 nm \cite{Nitrides_TiNRing}. The structure maintains its shape and optical performance after annealing at 1073 K,  whereas a Au metasurface with same ring structure was completely melted into nanoparticles under the same annealing temperature (Fig. \ref{fig:TiN} (f)).

\section{High-Temperature Metamaterials}

Since the scope of traditional optical materials, especially those with high thermal stability, is quite restricted, optical metamaterials with refractory building blocks are widely used in high-temperature thermal photonics \cite{MM_Refr1Heter,MM_Refr4}. These metamaterials can be engineered to have desired optical properties at some specific wavelength ranges \cite{MM_Book1,MM_Book2}, which may not be found in naturally occurring materials. 

~\begin{figure}[!ht]
    \centering
    \includegraphics[width=0.7\columnwidth]{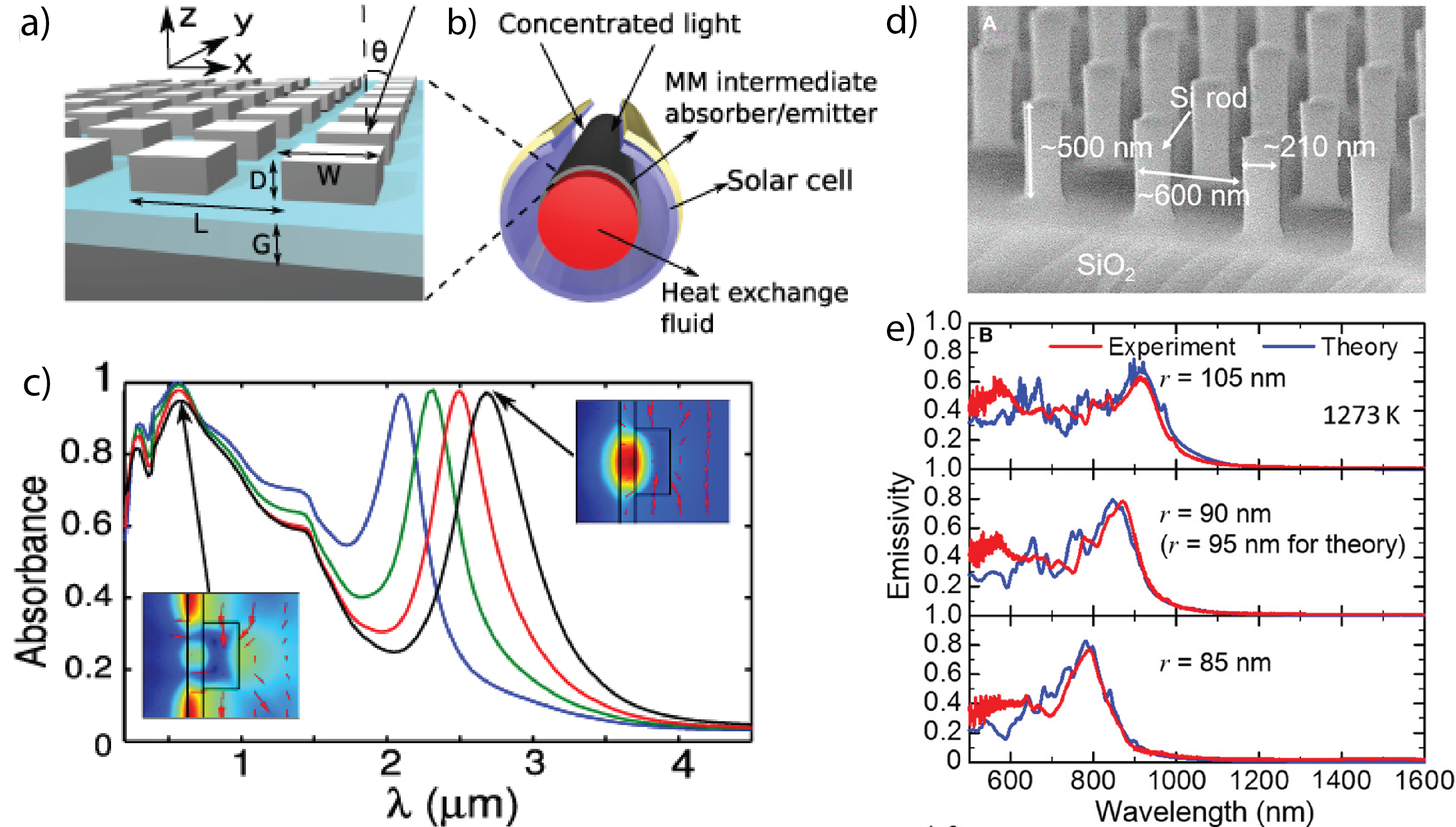}
    \caption{Engineering thermal radiation with refractory metamaterials. a) Schematic of a solar absorber/narrow-band thermal emitter (SANTE) structure. Gray regions: low reflectivity refractory metal (e.g., tungsten or molybdenum); light blue regions: dielectric spacer with high melting temperature (e.g., aluminum nitride). b) Conceptual schematic for a solar TPV system employing the SANTE structure. c) Calculated absorbance spectra of the SANTE structure for several geometric sizes. Absorbance peaks from left to right: [L, W, D, G] = [383, 206, 83, 27], [393, 230, 79, 24], [404, 252, 73, 23], and [396, 269, 72, 22] (in nm). Insets: field profiles for peak absorption frequencies. Surface: out-of-plane magnetic field. Arrows: in-plane electric field. Figure a, b, and c are adapted with permission from Ref. \cite{MM_SANTE}. Copyright 2012, IOP Publishing Ltd. d) SEM image of a fabricated refractory selective thermal emitter with Si rod arrays, r = 105 nm. e) Measured thermal emissivity of Si rod arrays (solid red lines) with rod radius of 105, 90, and 85 nm at 1273 K in surface normal direction. The theoretically calculated emissivity spectra are indicated by solid blue lines. Figure d and e are adapted from Ref. \cite{MM_SemiSirods}.}
    \label{fig:Meta}
\end{figure}

\subsection{Plasmonic/Polaritonic Metamaterials}

In an early example, Wu et al. proposed a large-area, nanoimprint-patterned metamaterial film (Fig. \ref{fig:Meta}(a))\cite{MM_SANTE}, which can be used as an integrated solar absorber/narrow-band thermal emitter (SANTE) in the solar TPV collection system shown in Fig. \ref{fig:Meta}(b). In this system, the metamaterial film is used to absorb the solar radiation that is concentrated on the cylindrical core, where heat exchange fluid is wrapped inside as a thermal reservoir. The solar energy is then re-emitted from the high-temperature metamaterial film into a spectral band that matches the bandgap of the coaxially positioned photovoltaic (PV) cell. To maximize the power conversion efficiency, both broad-band absorption of the solar radiation and narrow-band IR emission are desired for this SANTE film. A metamaterial design (Fig. \ref{fig:Meta}(a)) was proposed  to simultaneously achieve these two properties. The metamaterial is composed of a two-dimensional square array of high-loss or low-reflectivity refractory metal, a refractory conducting ground plane, and an ultra-thin spacer layer of high melting temperature dielectric separating the top patterned layer and bottom metal foil. The absorbance (emissivity) spectra of practical designs with tungsten squares, aluminum nitride spacers, and tungsten ground planes was calculated. In Fig. \ref{fig:Meta}(c), both the broad visible absorption bands and narrow IR emission peaks can be identified. With different geometric sizes, the IR emission peaks can also be tuned to match the bandgap of different PV cells. With this metamaterial-based SANTE film, calculations show that the power conversion efficiency would exceed the Shockley–Queisser limit for emitter temperatures above 1200 K. An efficiency as high as 41\% can be achieved at emitter temperatures of 2300 K. Costantini et al. fabricated a similar plasmonic metasurface with a tungsten (W) backplane, a SiN insulator layer, and sturdy platinum (Pt) square patches.\cite{MM_TETung} The emissivity spectrum of the fabricated device was measured at 600 $^\circ$C. Both high spectral selectivity and directionality have been observed. Wang et al. experimentally demonstrated a selective thermal emitter based on a bowtie-shaped SiC nanoantenna array.\cite{MM_Refr2} The origin of the narrowband thermal emission was studied using both near-field nanoimaging and far-field infrared spectroscopy. The extreme small modal volumes and the gap-dependence of thermal emission frequencies reveal that the SPhP resonances of SiC as well as the coupling between adjacent polaritonic nanostructures determine the observed thermal radiation characteristics of the metasurface.

\subsection{Semiconductor Metamaterials}

In the field of thermal metamaterials, not only the structural resonances but also the intrinsic properties of component materials have been exploited for engineering emissivity. In addition to the polaritonic and plasmonic resonances, in semiconductors, the interband transitions can also be used to engineer high-temperature thermal emission.  

For example, Asano et al. demonstrated a near-infrared–to–visible highly selective thermal emitter based on intrinsic Si meatamaterial \cite{MM_SemiSirods}. A scanning electron microscope (SEM) image presented in Fig. \ref{fig:Meta}(d) shows the fabricated sample with a Si rod array. Here, the thermal emission at low frequencies is suppressed because the electron thermal fluctuations are limited to those above the semiconductor bandgap. Meanwhile, thermal emission at the higher-frequency side of the spectrum, which is from the electronic resonance of interband transitions in Si, is further enhanced through the double-resonance of the Si array. The radius of the Si rods can be employed to tune the cut-off frequencies. The measured emissivities spectrum of the emitters at 1273 K are shown in Fig. \ref{fig:Meta}(e). High emissivities above 0.6 is observed below 1 $\mu$m. Wide-range emissivity measurements also indicate that these metamaterial emitters possess suppressed low emissivities up to 7000nm. Compared with some other metamaterial emitters using gold \cite{MM_TEGold1Antenna,MM_TEGold2,MM_TEGold4} and doped semiconductors \cite{MM_TESemi}, this design shows both higher thermal stability and a better suppression of longer-wavelength emission because of the use of intrinsic Si. It can be expected that, by using other semiconductors with wider bandgap, thermal emitters that work at even shorter wavelengths can be achieved by applying the same strategy.

\section{Epsilon-Near-Zero Materials}

ENZ frequency is the characteristic frequency where the dielectric permittivity of materials vanish. In natural materials, the ENZ response occurs at the bulk plasmon frequency of plasmonic metals and longitudinal optical phonon resonances of polaritonic dielectrics. Epsilon-near-zero media can be experimentally characterized by the presence of Ferrell-Berreman modes \cite{ENZ_FMode,ENZ_BMode,ENZ_FBMode}. Tuning the ENZ frequency of materials provides a unique approach to tailor high-temperature thermal emission. 

~\begin{figure}[!ht]
    \centering
    \includegraphics[width=0.7\columnwidth]{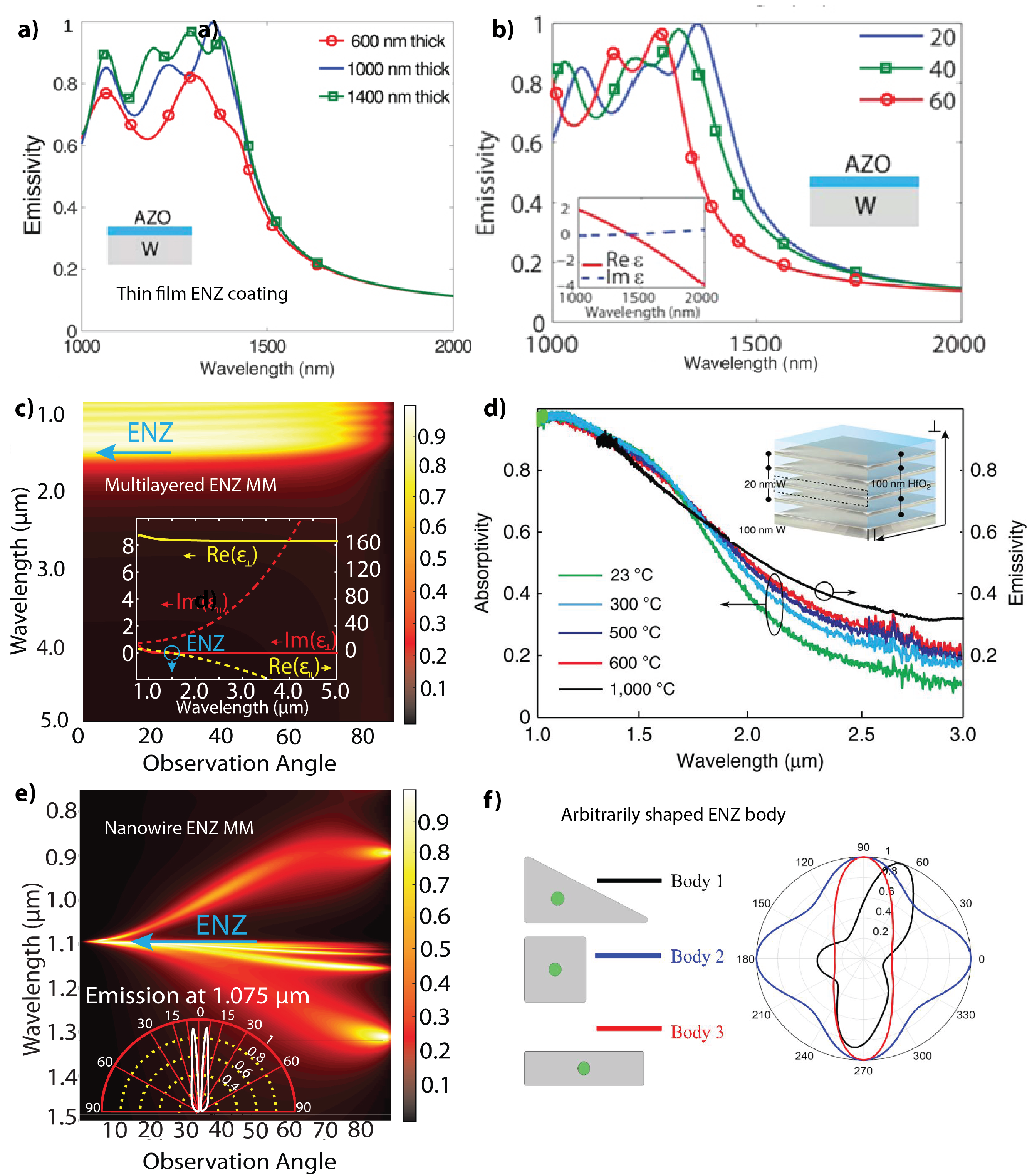}
    \caption{Radiative thermal properties of ENZ materials. a) Calculated emissivity spectra of an AZO thin film on a tungsten substrate at 20$^\circ$ emission angle and varying the AZO thickness. b) Calculated emissivity spectra of a 1000 nm AZO film on a tungsten substrate at different emission angles. Inset: real and imaginary part of the AZO dielectric permittivity. Figure a and b are adapted with permission from Ref. \cite{ENZ_Coating}. Copyright 2017, IOP Publishing Ltd. c) Calculated angularly-resolved  emissivity of a planar multilayered ENZ metamaterial. The metamaterial is composed of refractory materials tantalum and titanium dioxide. Inset: The effective medium parameters as functions of wavelength. Adapted with permission from Ref. \cite{ENZ_MMENP}. Copyright 2013, The Optical Society. d) Experimental observation of the OTT at high temperatures through normal incidence absorptivity measurements for the tungsten/hafnium dioxide ENZ metamaterial emitter at varying temperatures. Inset: Schematic of the fabricated device. Adapted from Ref. \cite{ENZ_MMExp}. e) P-polarized emissivity plot for a 450 nm thick metamaterial emitter consisting of a host matrix of aluminum oxide (Al2O3) embedded with 15 nm diameter silver nanowires in a 115 nm square unit cell using the effective medium approach. Inset: a polar plot of the emissivity at 1.075 $\mu$m. Adapted with permission from Ref. \cite{ENZ_MMENP}. Copyright 2013, The Optical Society. f) Sketch of the 2D SiC ENZ bodies augmented with Ge dielectric rods and the calculated emission pattern at its ENZ wavelength. Adapted from Ref. \cite{ENZ_Body}.}. 
    \label{fig:ENZ}
\end{figure}

\subsection{Plasmonic ENZ Thermal Emitter Coating}

Plasmonic materials always become reflective and exhibit a suppressed thermal radiation below the ENZ frequency. Thus, a thin film of plasmonic coating with engineered ENZ frequency can be employed to build selective thermal emitters. Pendharker et al. demonstrated a plasmonic thermal emitter coating (p-TECs) \cite{ENZ_Coating} with aluminum-doped zinc oxide (AZO, melting point 2200K \cite{ENZ_AZOMP}). The ENZ frequency is tuned to the near-infrared for the application of high temperature TPVs. The emissivity spectrum of AZO film on a tungsten substrate with varying film thickness (Fig. \ref{fig:ENZ}(a)) and at different emission angles (Fig. \ref{fig:ENZ}(b)) are numerically calculated. A sharp suppression in the emissivity above the ENZ wavelength of 1400 nm can be observed. This ENZ approach of thermal emission control is fundamentally different from other approaches using structural resonance, such as PhCs and anti-reflection coatings. The cut-off behavior is independent of the film thickness as well as the emission angle because the emission spectrum is dominated by the intrinsic material property. This thermal suppression of low energy photons provides an unique and robust solution to enhance the efficiency of TPV systems.

\subsection{Multilayered ENZ/Hyperbolic Metamaterials}

In naturally occurring materials, the possible range of ENZ frequencies is limited. Metamaterials with tunable effective permittivities can provide more freedom in designing ENZ thermal emitters. Due to the anisotropic nature of multi-layer metamaterials, ENZ behavior is generally associated with the optical topological transition (OTT) of the photonic isofrequency surface \cite{ENZ_OTT,ENZ_LJGuo}. On one side of the ENZ point, where both components of the effective permittivity are positive, an ellipsoidal isofrequency surface with bounded wavevectors is present. As a result, the medium supports radiative modes that can couple to free space, resulting in high absorptivity and emissivity.  On the other side of the ENZ point, one of the permittivity components turns negative, leading to hyperboloidal dispersion. An unbounded hyperboloid leads to modes with large wavevectors, which are momentum mismatched and cannot couple to free space. This results in a strongly suppressed absorptivity and emissivity in the corresponding spectral range.  

Molesky et al. proposed an ENZ thermal metamaterial with a planar multilayered structure \cite{ENZ_MMENP}. Here, refractory materials tantalum and titanium dioxide were used and structural parameters were carefully designed to obtain an ENZ wavelength at 1.5 $\mu$m. The thermal emission is suppressed below the ENZ frequency due to the high reflectivity of the structure, while the thermal radiation is enhanced due to the high absorption in the transparency window above the ENZ frequency (Fig. \ref{fig:ENZ}(c)). An experimental demonstration of such a multilayered ENZ emitter has been reported by Dyachenko et al. \cite{ENZ_MMExp}. The device is composed of layers of refractory metal W and transparent dielectric hafnium dioxide (HfO$_2$) on a W substrate. As shown in Fig. \ref{fig:ENZ}(d), absorption and emission measurements at temperatures up to 1273K confirm the long-wavelength thermal emission suppression above the ENZ point. Energy-dispersive X-ray spectroscopy (EDS) is also performed to investigate the thermal stability. Interlayer diffusion of the multilayered structure at higher temperatures is proven to be the primary mechanism of performance degradation. 

Note that the above works were related to far-field thermal emission. Investigation of the near-field thermal properties of such hyperbolic and epsilon-near-zero media using fluctuational electrodynamics showed that they exhibit topological transitions and broadband super-Planckian thermal emission in the near-field spectral energy density \cite{ENZ_NFSuper,ENZ_NFSuper2,ENZ_NFSuper3,ENZ_NFSuper4,ENZ_NFSuper5}.

\subsection{Nanowire ENZ Metamaterials}

The multilayered ENZ metamaterials produce mainly angularly independent (Fig. \ref{fig:ENZ}(c)) and polarization-independent thermal emission. Nanowire arrays are another alternative approach to realize ENZ metamaterials which can lead to highly directional emission patterns. The difference can be understood from the restricted motion of free electrons \cite{ENZ_MMENP}. In multilayered structures, nearly free electron motion is enabled in the parallel direction within the metal planes. In the perpendicular direction, the motion is largely confined due to the sub-wavelength thickness of the metal planes. Thus, a metal-like highly dispersive effective permittivity ($\epsilon_{\parallel}$) is produced in the parallel direction, while a dielectric-like permittivity ($\epsilon_{\perp}$) is present in the perpendicular direction. Above the ENZ frequency, the isofrequency surface becomes a hyperboloid with $\epsilon_{\parallel}<0$ and $\epsilon_{\perp}>0$, which is known as a type II hyperboloid \cite{BNNT_HyperbolicReview1,BNNT_BelovHyperbolic}. However, the confinement of electrons is exactly flipped in nanowire metamaterials, where free electrons are nearly propagating in the perpendicular direction along the axis of the nanowires. After the OTT, a type I hyperbolic metamaterial is formed with $\epsilon_{\parallel}>0$ and $\epsilon_{\perp}<0$. This fundamental difference between the two ENZ metamaterials thus leads to different emission patterns near the ENZ wavelength. Molesky et al. designed a metamaterial emitter consisting of Ag nanowire array embedded in an aluminum oxide (Al$_2$O$_3$) host \cite{ENZ_MMENP}. The angularly-resolved p-polarized emissivity spectrum is calculated using effective medium theory (Fig. \ref{fig:ENZ}(e)). Highly directive emission pattern and thus spatial-coherent thermal radiation are demonstrated in this structure. Note that another important characterstic of nanowire metamaterials as well as Mie scattering metamaterials is the epsilon-near-pole resonance frequency, where thermal emission can also be enhanced\cite{ENZ_MMENP,ENZ_RyanENP}.

\subsection{Arbitrarily Shaped ENZ Body}

Another strategy for engineering thermal emission is to utilize the spatially static fluctuating fields within ENZ bodies. Unlike other methodologies, where carefully designed geometry is needed to improve the spatial and temporal coherence, Liberal and Engheta theoretically proposed that enhanced spatial coherence would naturally occur on the surface of ENZ bodies because of the spatially static character of the temporally dynamic fields \cite{ENZ_Body}. 

As a case study, 2D SiC bodies with area $A=\lambda ^2$ are considered. SiC has a mid-infrared ENZ wavelength of 10.3 $\mu$m and a high quality factor ($\epsilon_{ENZ}=i0.03$). Circular germanium (Ge) rods with designed radii are introduced to investigate the effect of the resonant enhancement of constant fields. Ge is chosen because of the low loss and high dielectric constant around the ENZ wavelength. Thermal emission patterns of SiC ENZ bodies with and without the Ge rod are numerically calculated at the ENZ wavelengths. Directive thermal emission is revealed in both cases, which indicates the intrinsic spatial coherence of the ENZ media. Especially for the case when the dielectric rod is introduced, a constant field distribution on the external surface of the ENZ body is produced. As shown in Fig. \ref{fig:ENZ}(f), by changing the shape of the ENZ body, one can modify the lobes in the emission pattern, e.g. three lobes are observed in the triangle body (black), whereas only two lobes are present in the rectangle body, and four lobes are shown in the square body. Note that the thermal emission pattern is strikingly independent of the position of the dielectric rods because of the effectively enlarged wavelength and the spatially static fields inside ENZ bodies. Exploring this intrinsically enhanced spatial coherence and geometry-invariant character in refractory ENZ materials provides an appealing method for high-temperature thermal photonics. 

\section{Photonic Crystals}

PhCs are also widely used as a selective thermal emitter. The major physical feature of PhCs is the photonic bandgap, a frequency range where light propagation is forbidden \cite{PC_First1,PC_First2}. Thus, within the gap, very low photonic density of states are present, which translates into strongly suppressed thermal emission. Furthermore, as indicated by coupled-mode theory \cite{PC_CMT}, thermal emission of PhCs can be enhanced at designated wavelengths via Q-matching\cite{PC_QMatching}. Consequently, by adjusting the dimensions of PhCs, engineering of thermal emission spectra is achievable through the modification of photonic band structure and Q-matching condition.

~\begin{figure}[!ht]
    \centering
    \includegraphics[width=0.7\columnwidth]{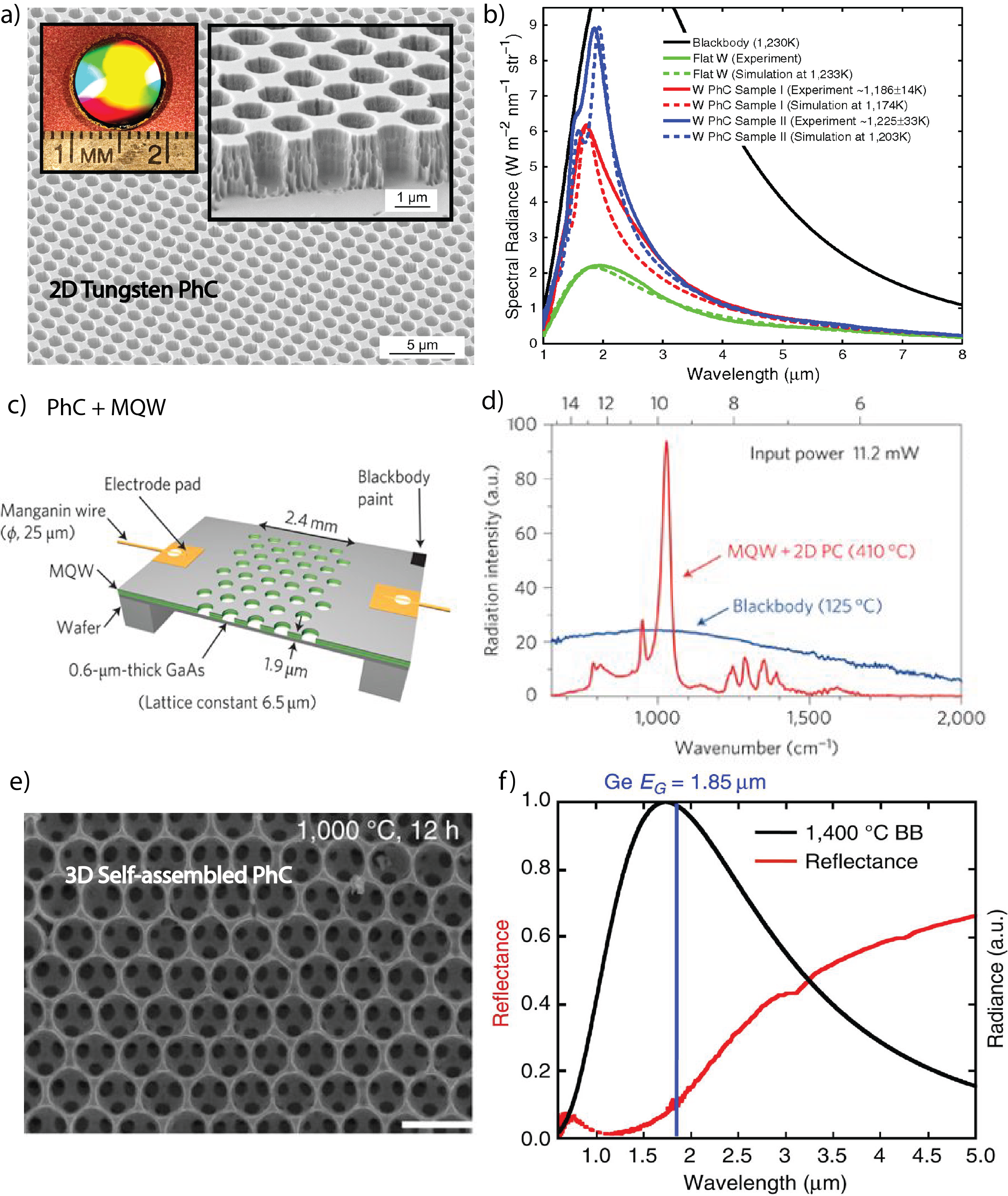}
    \caption{Engineering thermal emission with high temperature photonic crystals (PhCs). a) SEM Image of a fabricated tungsten (W) 2D PhC Sample over a large area, showing excellent uniformity. (Left Inset) Digital photo of the full 1-cm-diameter sample. (Right Inset) Magnified cross-sectional view of the W PhC sample. b) Measured and simulated normal spectral radiance of fabricated samples. The thermal emission data is obtained by electrically heating the sample to 1200 K in a vacuum chamber. Figure a and b are adapted from Ref. \cite{PC_2DCore}. c) Schematic of a hybrid thermal emitter, consisting of an MQW layer and a 2D PhC with a lattice constant of 6.5 $\mu$m. d) Thermal emission spectra of the MQW+PhC device and a blackbody control sample in the surface-normal direction under the same input power condition (11.2 mW). Figure c and d are adapted with permission from Ref. \cite{NodaMQW_NP}. Copyright 2012, Springer Nature. e) Top-view micrographs of a tungsten-inverse colloidal crystal after annealing at 1000$^\circ$C (1273 K) for 12h. Scale bar is 1 $\mu$m.  f) Reflection spectrum of the tungsten-inverse colloidal crystal, protected with HfO$_2$ and annealed at 1400$^\circ$C (1673 K) for 1h (red spectrum) plotted with the theoretical blackbody emission at 1400$^\circ$C (1673 K) (black spectrum). The blue line indicates the electronic band gap of germanium. Figure e and f are adapted with permission from Ref. \cite{PC_3DInverse3NC}. Copyright 2013, Springer Nature.}
    \label{fig:PhCs}
\end{figure}

\subsection{2D Photonic Crystals}

In recent years, metallic PhCs with large bandgaps and outstanding thermal stability have been intensely studied for high-temperature applications \cite{PC_Review}. For example, Yeng et at. designed and fabricated 2D metallic PhCs with single-crystal tungsten \cite{PC_2DCore}. The PhC is composed of a 2D square array of cylindrical air holes with a period $a$, fabricated on a W film through interferometric lithography. SEM images shown in Fig. \ref{fig:PhCs}(a) indicate that a long-range uniformity across a large area is achievable with the high-purity W. A digital photo shown in the left inset also indicates the high quality of the full 1-cm-diameter sample. 

In the 2D PhC design, each air hole can be considered to be a cylindrical cavity with radius r and depth d. A cutoff frequency is defined by the fundamental resonant frequency mode of the cavity, which is inversely proportional to the radius r. At a given value of r, the depth d is fixed to satisfy the Q-matching condition, where radiative and absorptive quality factors of the PhC cavity resonances need to be matched. Additionally, period $a$ is also carefully chosen to get rid of the diffractive modes and ensure sufficient sidewall thicknesses.

The experimentally measured emission spectra of the fabricated 2D PhCs are shown in Fig. \ref{fig:PhCs}(b). The data is obtained by electrically heating the sample to 1200 K in forming gas (5\% H$_2$-N) environment to prevent oxidation. Compared to the flat tungsten slab (green), the samples exhibit a near-blackbody emittance below the cut-off wavelengths as predicted by numerical simulation, while thermal emission above the cut-off is strongly suppressed at the same time. 

The thermal stability of such a W 2D microstructure was also examined through thermal cycling. After heating to 1200K and cooling down to room temperature several times, SEM and reflectance measurements show no noticeable change in the structure and optical properties of the sample are retained. This superior thermal stability, on the one hand, is due to the excellent ability of tungsten in withstanding high temperatures. On the other hand, it arises from the simplicity of the 2D  structure. Unlike the 1D PhC designs \cite{PC_1D1,PC_1D2}, where multilayer and multi-material structures are generally used, thermomechanical stress between layers and interfaces do not exist in this 2D PhC. Chemical reactions between constituent materials are also precluded since the cavities are just air holes. Note that similar 2D PhC structures have been demonstrated in other refractory materials, such as poly-crystalline tantalum \cite{PC_2DTa1,PC_2DTa2} and tantalum - tungsten alloy \cite{PC_2DTaW}. 

PhCs can also be made active with multiple quantum wells (MQWs), where the photonic resonance effects of PhCs are combined with the intersubband transitions (ISB-T) of MQWs to provide another degree of freedom of thermal emission control. A device demonstrated by Zoysa et al. is shown in Fig. \ref{fig:PhCs}(c). The device is based on  GaAs quantum wells and Al$_{0.3}$Ga$_{0.7}$As barriers on a GaAs substrate. The ISB-T wavelength is designed to be 9.7 $\mu$m. A 2D photonic-crystal consisting of a triangular lattice of air holes is fabricated into the MQWs and the resonant modes are carefully tuned to obtain the Q-matching condition. As shown in the thermal emission spectrum  (Fig. \ref{fig:PhCs}(d)), the device presents a strongly suppressed emission except for a sharp emission peak around 9.7 $\mu$m. Such a device thus demonstrates an efficient platform for thermal management: under the same input power from Joule heating, the device temperature and emission peak intensity can be much higher than that of a blackbody control sample.

\subsection{3D Photonic Crystals}

Three-dimensional PhCs can also be utilized to engineer thermal emission. They are more attractive compared with 2D PhCs because of the additional degree of freedom and controllable emission in all directions. Metallic and metal-coated 3D PhCs with woodpile structures have been fabricated and studied as selective thermal emitters \cite{PC_3DWoodPile1,PC_3DWoodPile2,PC_3DWoodPile3Coating}. However, their complex structures lead to a slow and costly fabrication process and also result in a slightly lower thermal stability because of the multiple interfaces. Thus, to address these challenges, inverse opal 3D PhCs fabricated with self-assembled templates were proposed and experimentally demonstrated \cite{PC_3DInverse0,PC_3DInverse1,PC_3DInverse2Tungsten,PC_3DInverse3NC,PC_3DInverse4}.

~\begin{figure}[!ht]
    \centering
    \includegraphics[width=1\columnwidth]{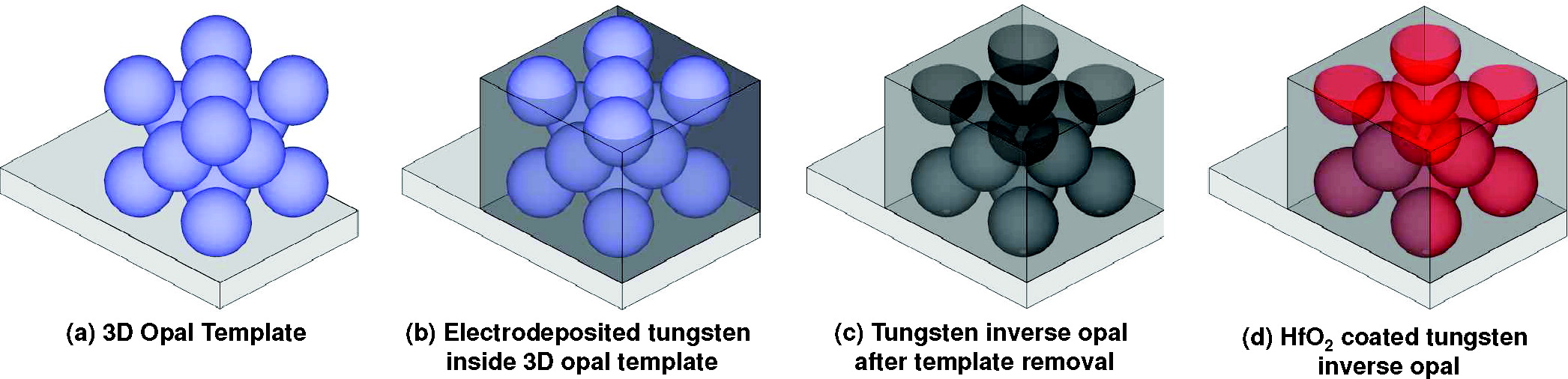}
    \caption{Fabrication process for an inverse opal 3D PhC with refractory coating. a) Silica opals are grown on tungsten foils. b) Tungsten is electrodeposited inside the 3D template. c) Tungsten inverse opals are obtained after template removal by HF etching. d) The tungsten inverse opals are coated with HfO$_2$ or Al$_2$O$_3$ by ALD to impart thermal stability. Adapted with permission from Ref. \cite{PC_3DInverse2Tungsten}. Copyright 2011, American Chemical Society.}
    \label{fig:Process}
\end{figure}

The fabrication process of an electrodeposited 3D tungsten PhCs with a refractory coating is shown in Fig. \ref{fig:Process} \cite{PC_3DInverse2Tungsten}. Here, silica opals are first grown on W foils (a). W is electrodeposited inside the 3D template to form the tungsten inverse opals (b). Then, the silica template is removed by hydrofluoric acid (HF) etching to obtain the 3D PhC (c). Finally, refractory coatings are introduced through atomic layer deposition (ALD) to improve thermal stability (d). SEM images reveal that the 3D structures are stable after annealed to 1000$^\circ$C for 12 hours (Fig. \ref{fig:PhCs}(e)). Arpin et al. also measured the reflection spectra of such 3D PhCs \cite{PC_3DInverse3NC}. As shown in Fig. \ref{fig:PhCs}(f), a suppressed reflection at shorter wavelengths indicates an enhanced thermal emission with photon energy higher than the bandgap of PV cells (germanium), which is very promising for TPV applications.  

\section{Nanotube Thermal Emitters}

~\begin{figure}[!ht]
    \centering
    \includegraphics[width=0.7\columnwidth]{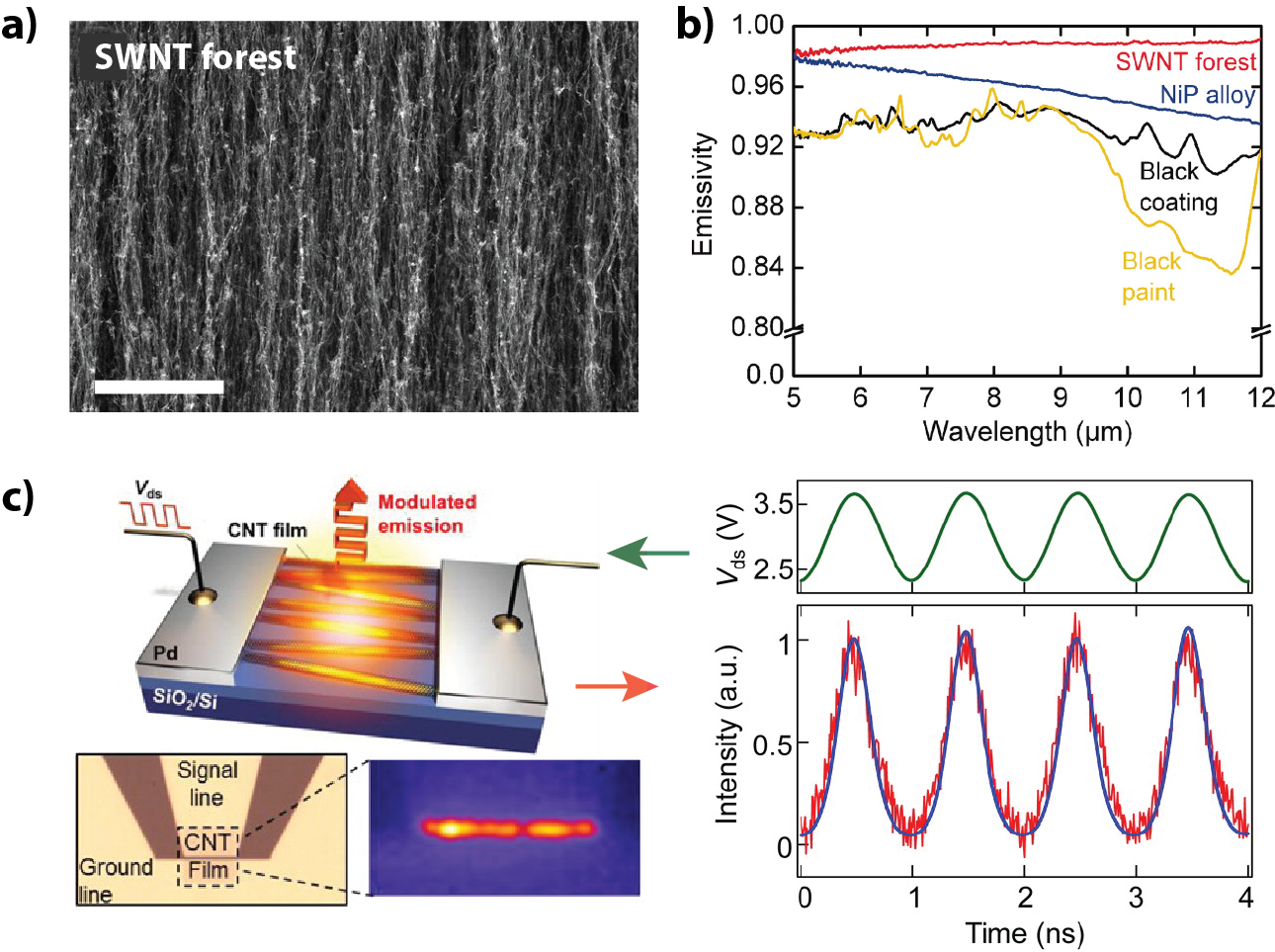}
    \caption{Engineering high-temperature thermal emission with nanotube systems. a) SEM image showing side surface of an SWNT forest (scale bar is 5 $\mu$m). b) Normal spectral emissivity of an SWNT forest (red line) is compared with commercially available black surfaces that are denoted as NiP alloy (blue), black coating (black), and black paint (yellow). Figure a and b are adapted from Ref. \cite{CNTBB_VASWCNT}. c) Left: Schematic of an electrically driven, ultrahigh-speed, on-chip thermal light emitter based on a carbon nanotube film. Right: Experimental high-speed thermal emission modulation (red curve) under a continuous drive of 1 GHz (green curve). A simulation based on the heat conduction equation (blue curve) agrees well with this experimental result. Adapted with permission from Ref. \cite{CNTEmitter_HighSpeed}. Copyright 2014, American Chemical Society. 
}
    \label{fig:nanotubes}
\end{figure}

\subsection{Blackbody Emitter}

A blackbody is an idealized object, on which all incident light will be absorbed. At thermal equilibrium, black bodies radiate isotropically and the spectrum can be predicted by Planck's law. Experimental realization of black bodies, especially those with high-temperature stability, is valuable since they are widely used as an infrared standard for temperature calibrations \cite{NodaMQW_NP}. 

CNTs are allotropes of carbon, including single-walled carbon nanotubes (SWCNTs) \cite{IijimaSWCNT} and multi-walled carbon nanotubes (MWCNTs)
 \cite{IijimaMWCNT}. They exist as tubes of graphite sheets with a length-to-diameter ratio greater than 1,000,000, for which they are generally considered as a one-dimensional material. Since first reported by Iijima in 1991 \cite{IijimaSWCNT}, CNTs have been intensely studied due to their novel mechanical, electrical, optical and thermal properties \cite{CNTReview1,CNTReview2,CNTReview3,CNTBooks}. 
CNTs have shown thermal stability at high temperatures \cite{CNTThermalStability}. They are also good candidate materials to make blackbody emitters. The blackbody behavior has been demonstrated in different CNT systems, including SWCNT bundles \cite{CNTBB_SWCNTBundle}, MWCNT bundles \cite{CNTBB_MWCNTBundle}, MWCNT films \cite{CNTBB_MWCNTFilms1,CNTBB_MWCNTFilms2}, individual semiconducting SWCNTs \cite{CNTBB_IndividualSWCNT} and individual MWCNTs \cite{CNTBB_IndividualMWCNT}. However, these blackbody emissions are polarization-dependent and generally manifest at some specific wavelength ranges.

To make an ideal blackbody emitter that works over a large spectral range, specific structures need to be introduced into CNT systems. For example, low-density vertically aligned CNT forests have been synthesized \cite{CNTBB_VACNTSyn1,CNTBB_VACNTSyn2} and proven to behave most similarly to black bodies \cite{CNTBB_VAMWCNT}. Mizuno et al. experimentally measured the blackbody behavior of vertically aligned SWCNT \cite{CNTBB_VASWCNT}. As shown in Fig. \ref{fig:nanotubes}(a), the normal spectral emissivity of the SWCNT forest (red) is more than 0.98 at 5-12 $\mu$m, which is significantly higher than the currently used black bodies, such as nickel-phosphorus (NiP) alloy (blue), black coating (black) and black paint (yellow) \cite{CNTBB_OtherBB}. In spectral ranges where thermal emissivity measurement is difficult, the reflection and transmission measurements show that the nanotube forest exhibits near-zero transmittance and extremely low (0.01–0.02) reflectance from 0.2 to 200 $\mu$m. The emissivity (absorptance) of the SWCNT forest is thus close to unity, which proves that the SWCNT forests are very close to an ideal blackbody emitter across an extremely wide spectral range from UV to far-infrared.

The blackbody behavior of CNT forests can be explained by its homogeneous sparseness and alignment \cite{CNTBB_VASWCNT}. On the one hand, in this system, CNTs account for about 3\% of the total volume, while the remaining 97\% of the volume is air. Therefore, the effective permittivity and refractive index of the forest are close to one \cite{CNTBB_EMT}. In this case, according to Fresnel's law, the reflectivity at the interface would be close to zero. On the other hand, due to the vertical alignment of CNTs in the forest, the near-normal incident light interacts with the CNTs mainly along the axial direction. Experimental and theoretical analysis has demonstrated that CNTs have strong optical anisotropy and that the light-interaction in the axial direction is very weak \cite{CNTBB_SWCNTPolarization}. Therefore, the incident light is hardly reflected or absorbed at the thin film top-surface. Most of the incident light proceeds into the SWCNT forest and is scattered by imperfectly aligned nanotubes. The light thus gets trapped inside the forest and is finally absorbed through a multiple scattering process. This high absorption leads to high thermal emission according to Kirchhoff's laws.

\subsection{High Speed On-Chip Emitter}

The 1D structure and high-temperature stability of CNTs can also be utilized to make on-chip thermal light sources \cite{CNTEmitter_IndividualSWCNT,CNTEmitter_Integrated,CNTEmitter_Display,CNTEmitter_Transistors}, which are of vital importance in optical communications and photonic circuits. Compared to conventional semiconductor materials, CNTs have a simple fabrication process and can be fabricated directly on silicon substrates, making them more compatible with existing and well-developed silicon-based optoelectronics. Additionally, the nanotube emitters are compact and can couple light directly into waveguides without the need of couplers \cite{CNTEmitter_HighSpeed}. CNTs also show low capacitance, low electromigration \cite{CNTEmitter_LowElectroMigra} and high thermal conductivity \cite{CNTEmitter_HighCond1,CNTEmitter_HighCond2,CNTEmitter_HighCond3,CNTEmitter_HighCond4}, which are essential to make high-speed light sources. As shown in Fig. \ref{fig:nanotubes}(b), Mori et al. demonstrated a high-speed modulation of thermal emission from  unsuspended SWCNTs film on an SiO2/Si substrate \cite{CNTEmitter_HighSpeed}. Due to the low heat capacity of the thin film and direct CNT–substrate contact, fast temperature modulation is enabled in this device. Under 1 GHz continuous input, the measured thermal emission signal is in good synchronization with the applied electric power. As shown at the bottom-right panel of Fig. \ref{fig:nanotubes}(b), experimentally measured thermal emission(red) is consistent with the simulation result (blue) based on the heat conduction equation of SWCNTs. Further simulations show that this electrically driven CNT thermal source can achieve a fast modulation up to 10 Gbps and generate short pulses down to 140 ps. Such CNT-based electrically driven high-speed on-chip light sources are highly valuable in high-density integrated photonics.

\subsection{Boron Nitride Nanotubes}

~\begin{figure}[!ht]
    \centering
    \includegraphics[width=0.8\columnwidth]{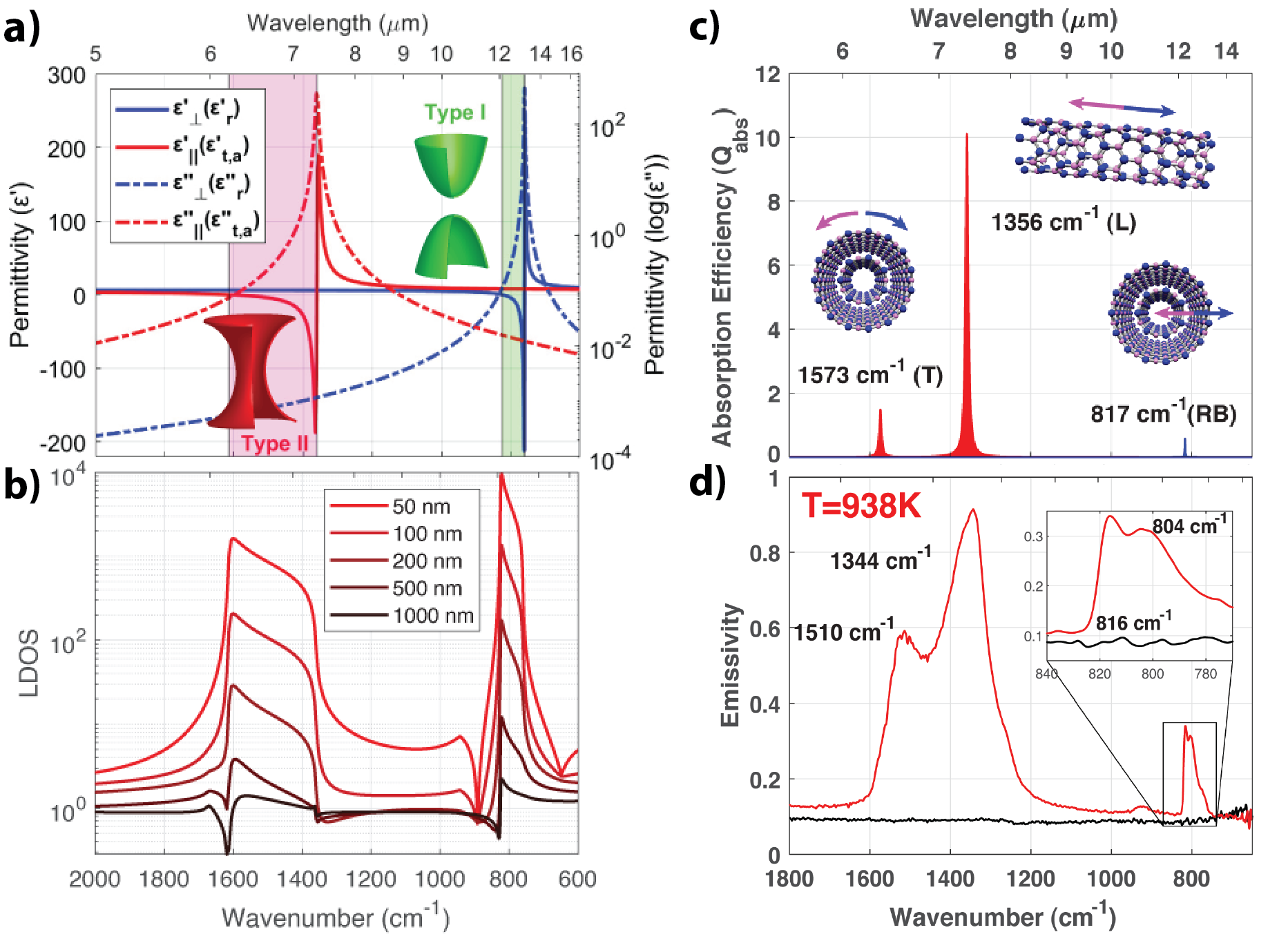}
    \caption{High temperature polaritons in  BNNTs. a) Real (solid) and imaginary (dashed) parts of the dielectric permittivity for hBN in the perpendicular (blue) and parallel (red) directions. The two Reststrahlen bands are highlighted in magenta (upper band) and green (lower band). b) LDOS above an hBN-vacuum interface at various distances from 50 nm (top) to 1000 nm (bottom). c) Absorption efficiency for an infinitely long cylindrical hBN nanotube. The T and L modes correspond to the upper frequency oscillator and are shown in red, while the RB mode corresponds to the lower frequency oscillator and is shown in blue. Schematics for each mode are shown and illustrate the boron (pink) and nitrogen (blue) ion movement for a given incident field. d) Spectral emissivity of the BNNT system on tungsten thin film (solid red) at 938 K. The emissivity spectrum of a tungsten thin film is shown (solid black) for reference. Adapted with permission from Ref. \cite{BNNT_Ryan} Copyright 2019, American Chemical Society.} 
    \label{fig:BNNT}
\end{figure}

Due to the 2D structure of hBN, the strong anisotropy results in different phonon resonance frequencies in perpendicular and parallel directions of 2D layers, which lead to two different components of permittivity as shown in Fig. \ref{fig:BNNT}(a). As both of these two components have two spectally distinct Reststrahlen Bands, hBN is a natural hyperbolic material with two different types of optical hyperbolicity. The unbounded nature of the hyperbolic isofrequency surface \cite{BNNT_HyperbolicReview1,BNNT_BelovHyperbolic,BNNT_ZubinHyperbolic} results in a high local density of states (LDOS) in the two hyperbolic Reststrahlen bands in the near-field of the hBN interface \cite{BNNT_LDOS} (Fig. \ref{fig:BNNT}(b)). By creating BNNTs out of hBN, the subwavelength structures of BNNTs can form natural thermal antennas, which couple these high LDOS near-field dark modes with far-field thermal radiation. Recently, the emissivity spectrum of disordered multi-walled BNNT thin films at high temperatures was demonstrated by Starko-Bowes et al. \cite{BNNT_Ryan}. Three different emission peaks are present at 1510, 1344 and 810 cm$^{-1}$, corresponding to the tangential(T), longitudinal(L), and radial buckling (RB) polaritonic modes of BNNTs \cite{BNNT_Modes1,BNNT_Modes2}. The emission spectrum is in good agreement with the theoretically calculated absorption spectrum based on Mie theory as illustrated in Fig. \ref{fig:BNNT}(c) and Fig. \ref{fig:BNNT}(d).

It was also noticed that the emission spectrum of the disordered systems still shows a certain angular dependence. As the emission angle is varied from normal direction to higher angles, the L mode emission peak is significantly suppressed, while the other two modes remain relatively unchanged. This antenna behavior comes from the quasi-ordered nature of BNNT thin-film systems, that is, most of the nanotubes are substantially lying down in the same plane, where the axis of nanotubes is parallel to the substrate. Because the L mode emission is from optical oscillations along the axial directions of the nanotubes, the radiation pattern of electric dipoles dictates that the radiation is strongest perpendicular to the nanotube axis and that the nanotubes do not radiate parallel to the nanotube axis. Simultaneously, the other two modes do not change with increasing angles due to the rotational symmetry. A calculation based on Mie theory also indicates that, for a vertically aligned BNNT forest, the L mode would completely disappear, and only the T and RB modes can be observed in the thermal emission spectrum.\cite{BNNT_Ryan} BNNTs behave as nano-antennas where the heat driven fluctuating dipole moments in optically active phonons lead to far-field thermal radiation.  

\section{Applications}

~\begin{figure}[!ht]
    \centering
    \includegraphics[width=0.7\columnwidth]{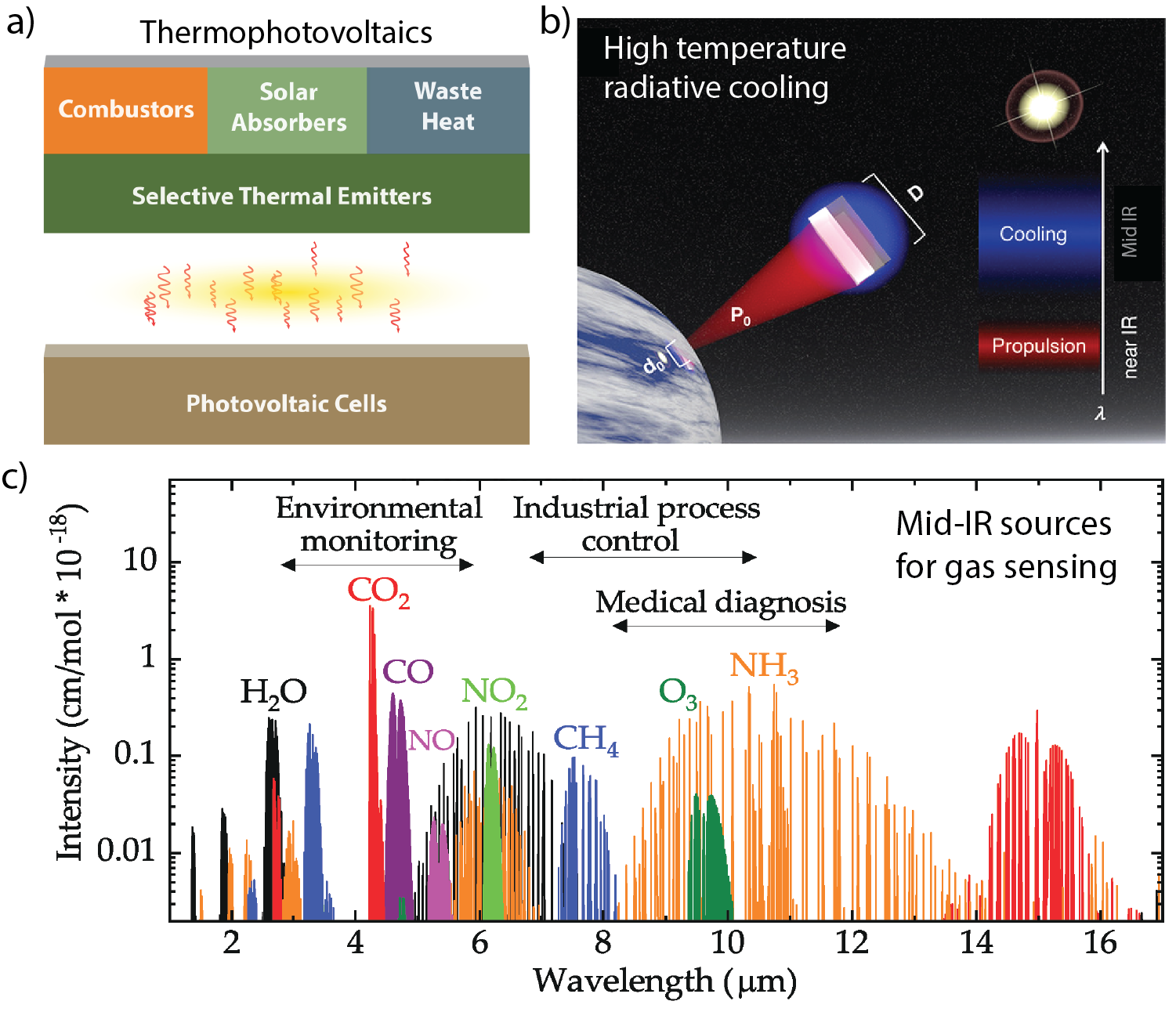}
    \caption{Applications of high temperature thermal photonics. a) Schematic illustration of typical TPV systems. b) Operating principle for propulsion and thermal management of a cosmic nanophotonic light sail. Radiation pressure from an Earth-based phased array propels the sail. For the light sail, multiband functionality is needed: low absorptivity in the (Doppler-broadened) spectral regime corresponding to the incident laser (red), and high emissivity in mid-IR for radiative cooling (blue). Adapted with permission from Ref. \cite{RC_LightSailCooling} Copyright 2018, American Chemical Society. c) Absorption spectra for selected gases in the mid IR region. High temperature mid-IR thermal sources can be used for gas sensing. Adapted from Ref. \cite{GasS_OptReview2}.}
    \label{fig:application}
\end{figure}

\subsection{Thermophotovoltaics}

TPV is one of the most important applications of high-temperature thermal photonics, which has received decades of intense studies \cite{TPV_Review1,TPV_Review2,TPV_Review3}.  As shown in Fig \ref{fig:application}a, a TPV system converts heat to electricity using as an intermediate emitter. The emitter operates at high temperatures and radiates light to the photovoltaic cell. Different approaches discussed above can be used to build selective thermal emitters to match the bandgap.

In solar energy harvesting, the maximum power conversion efficiency of conventional PV devices is found to be around 33\% for a single junction cell \cite{TPV_ShockleyLimitAccu}, because of the broadband nature of solar radiation and the radiative recombination in semiconductors. This efficiency limit is known as Shockley–Queisser limit \cite{TPV_ShockleyLimit}. Solar TPV can potentially overcome this limit with the help of broadband solar absorbers and selective thermal emitters \cite{TPV_Design1,TPV_Design2}. In the near-ideal case, where full solar concentration and near-monochromatic emission is assumed, Harder et al. theoretically demonstrated that the efficiency cap is as high as 85\cite{TPV_TPVEffiLimit}. Although numerous prototypes of solar TPV system have been experimentally achieved \cite{TPV_SExp3.2,TPV_SExp3.74,TPV_SExp6.2,TPV_SExp6.8,TPV_SExp8}, their power conversion efficiency is far from this ideal value due to many practical limitations.  

Early solar TPV systems were realized with solar concentrator, large absorption cavities and bulk material or rare earth compound emitters \cite{TPV_SBelow2,TPV_SExp0.18}, where the efficiency numbers were always lower than 1\%. Lenert et al. reported an solar TPV system with efficiency around 3.2\% \cite{TPV_SExp3.2}. A vertically aligned MWCNT forest was used as a near-perfect solar absorber and a 1D Si-SiO$_2$ PhC selective emitter were carefully designed to match the bandgap of a InGaAsSb PV cell. Additionally, by introducing a tandem plasma–interference optical filter, Bierman et al. demonstrated a higher conversion efficiency of 6.8\% \cite{TPV_SExp6.8}. 

TPV systems are also promising for applications that use other heat sources instead of solar radiation. For example, Chan et al. established an integrated power generator with a silicon propane micro combustor, a 1D Si-SiO$_2$ PhC selective thermal emitter and four 0.55 eV InGaAsSb PV cells \cite{TPV_ExpW2.5}. When the micro combustor operated at 1073K, an efficiency of 2.5\% was experimentally demonstrated. Very recently, Omair et al. demonstrated an ultra-efficient TPV system with a highly reflective rear mirror under the PV cell \cite{TPV_ExpW29.1}. The mirror maximizes the luminescence extraction of the PV cell and also helps reusing the low-energy thermal photons. A power conversion efficiency of 29.1\% is achieved with a Joule-heated graphite ribbon emitter at 1480 K. 

\subsection{High-Temperature Optical Coatings}
Optical coating is a well-developed technology that is used in various practical applications to obtain desired optical properties. Anti-reflection coatings are extensively used on eye-glasses and photographic lenses, while reflection coatings are commonly seen on the windows of vehicles and buildings. For high-temperature applications, by changing candidate materials to refractory ones, optical coatings also show its tremendous value in solar energy harvesting and radiative thermal engineering. 

For example, optical coatings are widely employed in solar absorbers, which are of crucial importance for applications including solar heating, solar thermoelectrics, and solar TPV. In these applications, high-temperature operations are generally required to achieve effective solar energy conversion. As a result, not only the thermal stability but also the thermal radiation loss need to be considered in the coating designs. Ideal solar absorbers require selective coatings that have high absorptivity in the solar spectral range (0.3-2.5$\mu$m) and low emissivity at longer wavelengths. Currently, large-scale commercial coatings employ black paints or cermet thin films. Although high solar absorptivity has been achieved, their overall energy conversion performance is still limited by their high thermal radiation losses\cite{Coatings_Commercial}. Moreover, as the fast development of concentrated solar power (CSP) technologies like parabolic trough collectors(PTC) or solar tower collectors(STC), higher operating temperatures of solar absorbers are desired, where current coatings have already fallen behind the requirement. Significant research progress has been made to optimize selective solar coatings by exploring different thermal photonic materials and structures. 

The multilayered stack of refractory materials is one of the most widely used coating designs. A typical design consists of a top anti-reflection (AR) layer, solar absorption layers in between, and an infrared-reflective layer at the bottom. Refractory oxide and nitride like Al$_2$O$_3$, SiO$_2$, TiO$_2$ , Y$_2$O$_3$, ZrO$_2$, HfN, TiN, Si$_3$N$_4$ are generally employed as the top AR layer, while IR reflective metals like W, Mo are used as the bottom layer\cite{Coatings_Materials}. Cermets materials, with Cr$_2$O$_3$, Al$_2$O$_3$, SiO$_2$, AlN as the matrix and Cr, Ni, W, Au, Ag, Cu, Mo as the metal inserts, are commonly found at the absorption layer. Many methods such as electroplating, anodization, physical and chemical vapor deposition, and solution-based fabrication have been utilized and further developed aiming for improved coating performance and large scale production\cite{Coatings_Cermet}.  

\subsection{High Temperature Radiative Cooling}

Thermal radiation is one of the fundamental routes through which a hot object spontaneously loses its energy. Enhanced radiative energy loss and suppressed energy absorption can result in a significant temperature decrease, which is known as radiative cooling \cite{RC_Review,RC_CGG}. Statistical studies have shown that, in the United States, nearly ten percent of energy consumption is for air conditioning \cite{RC_BuildingEnergy}. As a result, a completely passive approach such as radiative cooling offers a substantial impact on energy saving. 

Two basic principles need to be followed to realize efficient radiative cooling. The first is to reduce the energy absorption. As in daytime radiative cooling devices, an effective reflection of solar light is desired. The second is to enhance the radiative energy loss. The universe, with a temperature of 3 K \cite{RC_Universe}, provides an ideal heat sink. A transparency window in the atmosphere between 8 and 14 $\mu$m thus enables the cooling of emitters through a radiative heat transfer to the cold outer space. Note that the spectral range of the atmosphere window also aligns with the peak emission range of ambient-temperature objects. Near-room temperature radiative cooling has been experimentally demonstrated in different systems, like multilayered photonic strucutures \cite{RC_MultiLayer}, microspheres embedded polymer film \cite{RC_Scalable}, hierarchically porous polymer coatings \cite{RC_Porous},  and high-mechanical-strength cooling wood \cite{RC_Wood}.

Similar passive cooling approaches can also be exploited in high temperatures by replacing the conventional optical materials by refractory ones. For example, Fig. \ref{fig:application}(b) shows a proposed laser-propelled light sail system. An ultrathin, gram-sized light sail is illuminated by an Earth-based focused high-power laser (MW/cm$^2$), where the pressure of the light can propel the ultralight spacecraft to reach relativistic velocities \cite{RC_LightSail1}. In this application, the operating temperatures can potentially be higher than 1500K if the absorption of the propel laser is not fully suppressed. Thus the high-temperature stability of the sail needs to be ensured and effective cooling mechanisms need to be introduced because of the intense laser illumination. Conventional sources of cooling such as convection and conduction through a heat sink are not viable in space applications. Ilic et al. proposed a nanophotonic heterostructure design with multilayered Si-SiO$_2$ \cite{RC_LightSailCooling}. In this design, high reflection and thus low absorption of the near IR driven light is demonstrated for strong propulsion, while a high mid-infrared emission is simultaneously achieved for efficient radiative cooling. 

\subsection{High Temperature Thermal Sources}

Light sources are of vital importance in a variety of industrial and scientific applications. The thermal emission of high-temperature objects provides a simple and relatively efficient method to generate visible to mid-infrared light. However, traditional high-temperature thermal sources like incandescent light bulbs are incoherent. The emission is unpolarized, omnidirectional and spectrally broad, which largely limits their applications. Various thermal emitters have been experimentally demonstrated to engineer the properties of thermal emission according to desired applications as discussed in previous sections. Additionally, tailored thermal sources can also be achieved by engineering the photonic environment surrounding a plain thermal source \cite{Source_NN,Source_Fan,Source_Exclusion,Source_Broadband}. In this approach, photonic structures are carefully designed and brought close to thermal emitters. A higher operating temperature can be reached since the hot emitters don't need to be nano/micro-structured. For example, Ilic et al. demonstrated a high luminous efficiency incandescent light source by surrounding a plain tungsten filament with a nanophotonic interference system \cite{Source_NN}. The system can transmit visible light for a wide angular range and also reflect infrared light to reheat the filament. The estimated luminous efficiency of the light source is about 6.6\%, which is approaching commercial fluorescent bulbs and light emitting diodes.

~\begin{figure}[!ht]
    \centering
    \includegraphics[width=0.7\columnwidth]{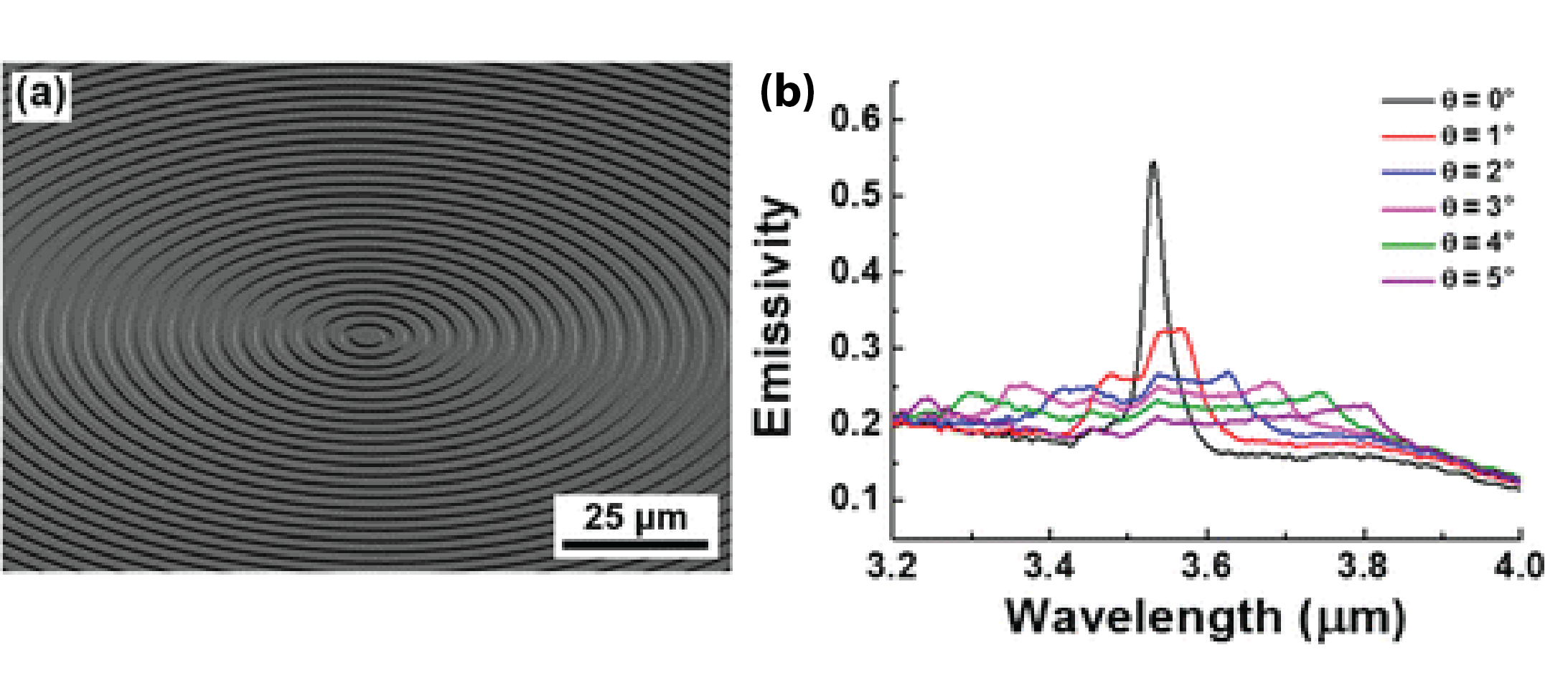}
    \caption{Thermal beaming from tungsten bullseye. a) SEM image of a W bullseye. The entire structure has a diameter of 2.1 mm and contains 300 circular concentric grooves.  b) Measured thermal-emissivity spectra of a W bullseye at various tilt angles ($\theta$) from normal at 900$^\circ$C.  Adapted with permission from Ref. \cite{Source_Bulleye} Copyright 2016, American Chemical Society.}
    \label{fig:Bulleye}
\end{figure}

\subsubsection{Thermal Beaming}

Highly directional thermal emission (i.e. thermal beaming) can be used to build novel light sources that have many practical applications. One typical design to achieve thermal beaming is the bullseye structure. In this structure, equally spaced circular concentric grooves act as gratings, coupling the thermally excited surface waves into monochromatic thermal emission normal to the surface. Park et al. first experimentally demonstrated this bullseye thermal emitter using tungsten (Fig. \ref{fig:Bulleye}(a)) \cite{Source_Bulleye}. The emissivity spectrum was measured at temperatures up to 900$^\circ$C. As shown in Fig. \ref{fig:Bulleye}(b), an angular divergence of 2$^\circ$ and a narrow spectral peak at normal direction were observed. The peak wavelength can be tuned by changing the groove periodicity and the thermal stability can also be further enhanced by introducing a thin layer of refractory coating.

\subsubsection{Gas Sensing} The detection of gases can be achieved utilizing the characteristic optical absorption of various gas species \cite{GasS_Review}. The gas concentration can be inferred from the attenuated light power via the Beer-Lambert Law. Compared with other gas sensing techniques, optical gas sensing offers a higher sensitivity and long-term stability \cite{GasS_OptReview1,GasS_OptReview2}. Because of the fundamental molecular vibration and rotation, most of the gases show fingerprint absorption lines or bands in the mid-infrared region. Fig. \ref{fig:application}(c) shows the mid-infrared absorption spectrum of various molecules together with their practical applications \cite{GasS_OptReview2}. 

One of the most important components in the optical gas sensor technique is light sources with emission corresponding to the absorption fingerprint of target gases. Here, thermal emission can be used as highly efficient sources due to the broadband spectrum. High-temperature operation is desired to get sufficient source signal and thus an appropriate signal to noise ratio (SNR). Especially for those gases with short-wavelength absorption lines, such as H$_2$O, higher temperature is required for enough source light intensity, as indicated by Planck's law and Wien's displacement law. For an individual gas species, selective thermal emitters with a narrow emission band can be built using the principles discussed in previous sections. For example, Moelders et al. developed a MEMS CO2 sensor with a PhC selective emitters \cite{GasS_EA1}. A detection limit of ~1600 ppm CO2 has been achieved. Lochbaum et at. demonstrated a gas sensing system with an on-chip narrow band thermal source by integrating MEMS heaters with Al$_2$O$_3$ metamaterial emitters \cite{GasS_EA2}. These low-cost and compact gas sensing systems are well suited for the next-generation medical and environmental applications.  

\subsection{Noisy Thermal Devices}
Thermal management in modern nanoelectronics is crucial due to the deleterious effects of on-chip heat dissipation \cite{cahill}. To harness this excess heat readily available at the nanoscale, apart from the TPV technology, new information processing functionalities based on heat (rather than electric current) have been explored within the last decade. In the context of using radiative heat (thermal photons), this includes many theoretical and experimental works which have demonstrated interesting functionalities designed to operate over a wide temperature range ($T \sim 300$K-$1500$K). This includes thermal diodes~\cite{otey2010thermal,fiorino2018thermal}, transistors~\cite{abdallah2014near,joulain2016quantum}, memories~\cite{khandekar2017thermal,elzouka2014near,morsy2019high} and logic gates~\cite{abdallah2016boolean}. Innovation of such thermal analogues of electronic devices can be particularly useful in extreme environments and at high temperatures where the performance of traditional Si-based electronics is degraded. 

\section{Conclusion}
New frontiers are emerging in the field of thermal photonics. They include the control of properties beyond spectrum and directionality such as the spin angular momentum of thermal radiation\cite{ChinmaySpin2}. Non-reciprocity and magnetic field based thermal emission control is another open area for the field \cite{FanNonRecip,ChinmaySpin2}. In addition, we anticipate the inevitable convergence of photonic heat radiation with conventional conduction and convection. As already shown in the area of porous materials and nanofluids, accurate evaluation and engineering of effective thermal transport need a comprehensive and simultaneous consideration of all three thermal transport mechanisms\cite{Conclusion_Porous, Conclusion_Nanofluids}. These convergences will lead to large macro-scale multi-physics tools and devices with huge industrial impact. 

\clearpage
\subsubsection{Acknowledgements}
This material is based upon work supported by the U.S. Department of Energy, Office of Basic Energy Sciences under award number DE-SC0017717.

%
%
%
\bibliographystyle{bib_ijuq.bst}
\bibliography{biblio}
\end{document}